\documentclass[10pt]{article}
\usepackage{epsfig,amsmath,float,fullpage}%,chicago}
\usepackage{wrapfig}
\usepackage{url}
\usepackage{sidecap}
\usepackage{multirow}
\usepackage{amsmath}
\usepackage{graphicx}
\usepackage{amsthm}
\usepackage{amssymb}
\usepackage{natbib}
\usepackage{setspace}

\newcommand{\ds}{\displaystyle}
\newcommand{\mrm}{\mathrm}

\usepackage{color}

\begin{document}

%\title{Long-term Analysis of Androgen Deprivation Therapy in Prostate Cancer Biochemical Failure}
\title{Modeling Long-term Outcomes and Treatment Effects After Androgen Deprivation Therapy for Prostate Cancer}

\renewcommand{\arraystretch}{0.65}
\author{\small Yolanda Hagar, James J. Dignam, and Vanja Dukic\footnote{
\textcolor{black}{Yolanda Hagar  is a postdoctoral researcher in Applied Mathematics, University of Colorado at Boulder.  James Dignam is an Associate Professor, Department of  Health Studies, University of Chicago. Vanja Dukic is an Associate Professor in Applied Mathematics, University of Colorado at Boulder.   
The authors would like to acknowledge National Institute of Health grants R21 DA027624-01 and U01 GM087729-03, and the National Science Foundation grants NSF-DEB 1316334 and NSF-GEO 1211668 that partially supported Dr. Dukic and Dr. Hagar. National Institutes of Health grants U10 CA21661 and  U10 CA37422 (from the  National Cancer Institute) supported conduct of the clinical trial and Dr. Dignam's work. Additional support was provided by Pennsylvania Department of Health 2011 Formula Grant (the Department specifically declaims responsibility for any analyses, interpretations or conclusions.)  The project utilized the Janus supercomputer, which is supported by the National Science Foundation (award number CNS-0821794) and the University of Colorado - Boulder. The Janus supercomputer is a joint effort of the University of Colorado - Boulder, the University of Colorado - Denver, and the National Center for Atmospheric Research. Janus is operated by the University of Colorado - Boulder.
Correspondence emails: {\tt yolanda.hagar@colorado.edu,  vanja.dukic@colorado.edu.}}
}}

\date{  }
\maketitle

\thispagestyle{empty}
\baselineskip 12pt

\begin{abstract}
% 80-200 words
\noindent
Analyzing outcomes in long-term cancer survivor studies can be complex.  The effects of predictors on the failure process may be difficult to assess over longer periods of time, as the commonly used assumption of proportionality of hazards holding over an extended period is often questionable. In this manuscript, we compare seven different survival models that estimate the hazard rate and the effects of proportional and non-proportional covariates.  In particular, we focus on an extension of the the multi-resolution hazard (MRH) estimator, combining a non-proportional hierarchical MRH approach with a data-driven pruning algorithm that allows for computational efficiency and produces robust estimates even in times of few observed failures.  Using data from a large-scale randomized prostate cancer clinical trial, we examine patterns of biochemical failure and estimate the time-varying effects of androgen deprivation therapy treatment and other covariates.  We compare the impact of different modeling strategies and smoothness assumptions on the estimated treatment effect.  Our results  show that the benefits of treatment diminish over time, possibly with implications for future treatment protocols.\\

\end{abstract}

\newpage

%%%%%%%%%%%%%%%%%%%%%%%%%%%%%%%%%%%%%%%%%%%%% %%%%%%%%%%%%%%%%%%%%%%  
%%INTRODUCTION 
%%%%%%%%%%%%%%%%%%%%%%%%%%%%%%%%%%%%%%%%%%%%% %%%%%%%%%%%%%%%%%%%%%%  

\section{Introduction}
Many human cancers today are  considered ``chronic diseases'', with long-term disease trajectories and multiple co-morbidities. Consequently, long-term cancer outcomes may be affected by numerous factors, ranging from obvious patient and treatment characteristics to secular and health care trends that  affect treatment policy and practice.  However, the long-term nature of patient and health-care related processes and changing complexity of information can make the analysis of long-term patterns in cancer survivor data sets  challenging.  In addition to  sparsely observed failure times, these data often exhibit non-proportional effects over time, requiring flexible and computationally efficient statistical methods to characterize the evolving failure hazard.

The motivating problem in this  article comes from a set of prostate cancer clinical trials from the Radiation Therapy Oncology Group (RTOG), which are specifically designed to examine the effects of the length of androgen deprivation (AD) therapy on  disease-free and overall survival time. As a result of the insight gained from these studies over the years, the short-term benefits of AD therapy have been well-understood to delay the time until prostate cancer recurrence and until death \citep{ Pilepich2, Pilepich1}. In addition,  longer-duration AD therapy has proven more beneficial than a shorter-duration therapy \citep{Horwitz}. However, given that AD treatments can have unpleasant side-effects,  clinicians have been reluctant to assign androgen deprivation for longer than necessary. 

Many questions remain regarding the relationship between the length of AD therapy and long-term outcomes, such as eventual time to  recurrence \citep{Schroder}.   These open questions still exist in part because prostate cancer is generally a slow cancer to progress \citep{prostateSlow}. While prostate cancer patients tend to survive longer  and are thus  observed over extended periods of time, treatment benefits have been difficult to precisely ascertain due to a multitude of co-morbidities and sparsity of information at long follow-up times.   

Thus, assessing the long-term benefits of different duration of AD therapy for patients in different risk classes would be of great value  to clinical practice and management of prostate cancer patients in general \citep{Schroder}.  Gaining insight into recurrence patterns and quantifying the degree and length of long-term benefits over time would greatly improve the quality of life for men with this disease.  For this reason, the focus of our paper goes beyond integrated summaries such as survival curves and cumulative incidence functions, and concentrates on estimation and inference about the time-dynamic hazard function in the presence of covariates, and the time-evolving predictive probabilities of disease recurrence.

The underlying statistical approach employed in this paper is an extension of  the multi-resolution hazard (MRH) model, a Bayesian semi-parametric hazard rate estimator previously presented and used in \cite{Bouman, Bouman2}, \cite{Dukic}, \cite{Dignam},  \cite{Yprune}, and \cite{sadm}.  This flexible class of models for time-to-event data is based on the Polya tree methodology, and is also similar in spirit to the adaptive piece-wise constant exponential models \citep{Ibrahim}.  The MRH parametrization is designed for multi-resolution inference capable of accommodating periods of sparse events and varying smoothness, typical in long-term studies. In addition, the MRH model accommodates both  proportional and non-proportional effects of predictors over time.  The current methodology employs the pruning algorithm presented in \cite{Yprune}, which performs adaptive and data-driven ``pre-smoothing'' of the hazard rate, via merging of time intervals with similar hazard levels.  Pruning has been shown to increase computational efficiency and reduce overall uncertainty in hazard rate estimation  in the presence of periods with smooth hazard rate and low event counts \citep{Yprune}.  All MRH models have been fitted  using the `MRH'' R package \citep{MRHR}.

This paper is organized as follows. In the next section, we provide a short overview of the prostate cancer clinical trials data, and the statistical issues related to these studies with long-term follow-up.  Sections 3 and 4  present the corresponding MRH methodology and implementation. Section 5 presents the analysis of biochemical failure in prostate cancer, with comparisons of the MRH approach to a set of alternative models: the Cox proportional hazards model, an extended Cox model that includes a time-varying treatment effect, a non-proportional hazards Weibull parametric model,  a semi-parametric Bayesian accelerated failure time model, a dependent Dirichlet Process survival model, and two piece-wise exponential models.  In addition we perform a sensitivity analysis to the priors in the MRH model.  The article concludes with a discussion of the clinical and statistical importance and implications  of our findings.  

\section{Motivating Example: Androgen Deprivation in Prostate Cancer}\label{sec:dataDescr}

Typical prostate treatment involves radiation therapy combined with some form and duration of hormone treatment, which is known as androgen deprivation (AD) therapy. The motivation for the current analysis is the characterization of the hazard rate of time-to-biochemical failure, adjusted for the (potentially)  time-varying effects of AD therapy and several key covariates, as described below.  

The outcome of interest in this analysis,  ``biochemical failure'', is defined according to the Phoenix definition as a two-unit rise in prostate specific antigen (PSA) level following a post-treatment PSA nadir \citep{defineBF}. Prostate specific antigen is a glycoprotein produced almost solely by prostatic epithelial cells, and is a biomarker routinely measured to screen for possible presence of prostate cancer.  Men with prostatic diseases (including cancer) can have high serum PSA levels due to structural changes in the prostate gland as well as to the enhanced production of PSA; therefore, elevated levels have long been used as a possible indication of the presence of prostate cancer, including residual or recurrent disease after treatment \citep{PSAorig3, PSAorig2, PSAorig1}.  Although recent studies question PSA as a screening method for initial prostate cancer diagnoses \citep{PSAcontr1,PSAcontr2,PSAcontr3}, the examination of the rise in PSA levels post-cancer treatment is still considered by many to be a useful clinical tool for assessing the risk or presence of prostate cancer recurrence. 

The rises in PSA levels can lead to what is termed ``biochemical failure", which in itself is not currently considered a clinical endpoint. However, biochemical failure is thought to importantly portend advancing (and possibly sub-clinical) disease.   Prostate cancer mortality risk might also be affected by patterns in biochemical failures \citep{BFuseful2,BFuseful3,BFuseful1} over time. However, because of its lack of direct clinical consequences, its use as a primary endpoint in clinical trials has been controversial. 

Nonetheless, characterization of the biochemical failure hazard over time, particularly within different  patient subgroups defined by disease characteristics or treatment regimens, would provide a strong foundation for determining how this endpoint may relate to the levels of risk for clinical recurrence and death.  A better understanding of these recurrence patterns  over time could be of great value for clinical management, design of clinical trials, and biologic insights into prostate cancer progression in different population subgroups.

The data we use to analyze biochemical failure hazard come from the Radiation Therapy Oncology Group (RTOG), which is a national clinical cooperative group that has been funded by the National Cancer Institute (NCI) since 1968 in an effort to increase survival and improve the quality of life for cancer patients.  The group consists of both clinical and laboratory investigators from over 360 institutions across the United States and Canada and includes in its membership nearly 90\% of all NCI-designated comprehensive and clinical cancer centers. The specific RTOG clinical trial we examine in this paper is RTOG 92-02, which is part of a series of RTOG clinical trials conducted from the 1980s to the present. These rich studies provide a wealth of data sources for studying the ``natural history" of prostate cancer as it is presently defined and managed clinically.  

RTOG 92-02 was a multi-center study, designed with the primary objective of evaluating the effectiveness of androgen deprivation therapy on prostate cancer disease progression and survival.  Between 1992 and 1995, 1,521 participants with locally advanced high risk prostate cancer were accrued in over 200 treatment centers across the country. During the trial, all patients received  4 months of androgen deprivation (AD) therapy including goserelin and flutamide, in addition to  external beam radiation therapy. Subjects were  then randomized to either no further AD therapy (the ``+0m AD group''), or an additional 24 months of goserelin (the ``+24m AD group'') using the treatment allocation scheme described by \cite{zelen}, and were stratified according to stage, pretreatment PSA, grade, and nodal status.  Given RTOG's   long history of high quality, well-randomized clinical trials with strictly executed protocols in  each institution \citep{rtogwebsite}, analyses of the pooled data (without considering center heterogeneity) have been dominant in previous analyses of these trials (for example, see  \cite{pooledex3}, \cite{pooledex2}, \cite{Horwitz}, and \cite{pooledex1}), and our analysis here follows suit. Further protocol details and study description can also be found in \cite{Horwitz} and \cite{rtogwebsite9202}. 

For each patient enrolled,  several measures of aggressiveness and severity of the original cancer were recorded at baseline: the Gleason score,  T-stage of the tumor, and the PSA level at diagnosis. The Gleason score is assigned by a pathologist after microscopic examination of a tumor biopsy.  Based on the degree to which the prostate cells have become altered, a Gleason score ranging from 2-10 is assigned, with scores between 2 and 4 indicating almost normal cells that pose little danger, and scores above 8 indicating very abnormal cells and a cancer that could be  aggressive \citep{gleason}.  

The American Joint Committee on Cancer (AJCC) staging criteria is used to assign the tumor a T-stage, which indicates the extent that the primary tumor has spread. (In this analysis, we omit the `N' and `M' components of AJCC staging, as they are only applicable to non-localized cancer cases).  Because all patients in the RTOG 92-02 trial were selected as ``high-risk'' by pre-specified criteria \citep{rtogwebsite9202}, our data set only contains men with tumors of stage two (T2) through four (T4).  Stage T2 indicates that the tumor can be felt during a physical examination, but has not spread outside the prostate, stage T3 indicates that the tumor has spread throughout the prostate (or the ``prostatic capsule"), and stage T4 indicates the tumor has spread beyond the prostate \citep{tstage1, tstage2}.  

In conjunction with the Gleason score and the T-stage, PSA levels at the time of diagnosis are an important component of prostate cancer staging, with very high levels frequently thought to be associated with a more severe form of prostate cancer.   Since the Gleason score, T-stage, and PSA level are all important components of the cancer  severity at the time of diagnosis, they, in addition to the age at diagnosis, will be considered as predictors in the   biochemical failure analysis.

The final data set considered  in this analysis comprises 1,421 subjects, after the removal of 100 subjects with  missing Gleason scores. Of those 1,421 subjects, 705 men (49.6\%) received no additional AD therapy (were placed in the ``+0m AD therapy'' group) and 716 men (50.4\%) received additional 24 months of AD therapy (the ``+24m AD therapy'' group).  The sample median time to biochemical failure is 4.9 years (SD = 3.9, range = 0.03 - 13.65). Biochemical failure was observed for 50.4\% of the patients before the end of the study period.  The sample median age at baseline was 70 years  (SD = 6.5, range = 43-88).   Table \ref{tab:summcateg} summarizes the sample  characteristics in more detail.
\begin{table}
	\caption{Sample characteristics of 1,421 patients in the final RTOG 92-02 trial data set, by treatment group, with age at diagnosis in 10-year increments, Gleason scores categorized by grade, and T-stage categorized into levels 2-4.}\label{tab:summcateg}
\centering
\begin{tabular}{|cc|cc|cc|cc|}
	\hline
	\hline
	&&&&&&&\\[.01ex]
	& &\multicolumn{2}{|c|}{+0m AD} & \multicolumn{2}{|c|}{+24m AD} &\multicolumn{2}{|c|}{Total Sample}\\[.01ex]
	\hline
   	&& N& \% & N&  \%  & N&  \%      \\[.01ex]
	\hline
	\hline
	&&&&&&&\\[.01ex]
	Patients &&705&49.6&716&50.4&1421&100.0\\[.5ex]
	\hline
	\hline
	&&&&&&&\\[.01ex]
	\multirow{4}{*}{Age}&Less than 60 years&44 &3.1&51&3.6 &95&6.7\\
	&60 to 70- years& 274&19.3& 260&18.3&534&37.6\\
	&70 to 80- years&363 &25.5&371 &26.1&734&51.7\\
	&80 or more years& 24&1.7&34&2.4 &58&4.1\\[.01ex]
	\hline
	\hline
	&&&&&&&\\[.01ex]
	\multirow{3}{*}{Gleason score}&Low grade (2-4) &56&3.9& 47&3.3&103&7.2\\
	&Intermediate grade (5-7) &462&32.5 &495&34.8&957&67.3\\
	&High grade (8-10)&187 &13.2&174&12.2 &361&25.4\\[.01ex]
	\hline
	\hline
	&&&&&&&\\[.01ex]
	\multirow{3}{*}{T-stage}&T2& 325&22.9&331&23.3 &656&46.2\\
	&T3&360& 25.3&353& 24.8&713&50.2\\
	&T4&20& 1.4&32& 2.3&52&3.7\\[.01ex]
	\hline
	\hline
\end{tabular}
\end{table}

\section{Multi-resolution Hazard Model}\label{sec:MRH}
The hazard function of the time to biochemical failure $T$ is defined as
$h(t) = \lim_{\Delta \to 0} {P(t \le T < t+\Delta \mid T \ge t)}/{\Delta} = {f(t)}/{S(t)},$
where $S(t) = P(T > t)$ is the survival function and $f(t) = -S'(t)$ is the probability density function of $T$. 
While the hazard rate can provide a more detailed pattern over time that is not always visible in aggregate measures such as the survival curve or cumulative hazard function, it may  also be more difficult to  estimate reliably. This is  particularly true if event counts  are  sparse, as is often the case in studies that follow subjects for extended periods of time.  

Various statistical estimators have been developed for the hazard function \citep{Andersen}. They vary from classic parametric methods that assume a known family of failure time distributions such as Exponential or Weibull, to semi- or non-parametric smoothing methods such as those in \cite{GrayBiom, Gray, graycompstat}, \cite{Hess}, and \cite{Sargent},  to methods using process priors in the context of non-parametric Bayesian hazard estimation. In this latter group, \cite{Hjort} introduces the beta process prior and \cite{leekim} develop a computational algorithm to approximate a beta process by generating a sample path from a compound Poisson process. Similarly, \cite{kalb} and  \cite{bur} model the cumulative hazard function as a gamma process. Correlated process priors, such as those used by \cite{arjas}, rely on a martingale jump process to model the hazard rate. In \cite{nieto}, the  prior correlation is  introduced via a latent Poisson process between two adjacent hazard increments. Other non-parametric Bayes hazard rate estimation methods are reviewed in \cite{sinha97} and \cite{mullerbook}.  

Different predictors and covariates can be included in hazard modeling under the proportional hazards assumption \citep{CoxPH,Ibrahim,Bouman2}.  However, with longer-term follow-up, the assumption of constant proportionality between the hazard rates for different patient subgroups may be questionable, either because the effects change throughout the course of a study, or because the remaining patients constitute a subpopulation significantly distinct from the population at baseline.  In these instances, it is important to relax the proportionality of hazards assumption over time. 

One of the simplest ways to accommodate non-proportionality of hazard functions among groups of patients is to perform a stratified analysis and estimate each group's hazard function separately.  However, this simple method cannot jointly estimate the hazard rates and effects of predictors, nor allow for correct quantification of  uncertainty.  More sophisticated investigations involve the use of piece-wise hazard functions, examples of which can be found in \cite{Holford76, Holford80}, \cite{Laird}, and \cite{Taulbee}.  Other methods address non-proportionality issues by pre-testing and comparison of two survival or hazard functions through graphics and asymptotic confidence bands \citep{Dabrowska,Parzen,McKeague}, or through asymptotic confidence bands for changes in the predictor effects over time \citep{Wei, Dong}.  Some of these methods mentioned require large sample sizes for inference, and their performance can degrade over time in studies with sparser outcomes in later periods.  Alternative approaches to handling non-proportionality have been implemented through accelerated failure time (AFT) models (with initial work done by \cite{AFTbuckley} and a thorough review found in \cite{KalbPrenticeBook}), which accommodate non-proportionality through specific parameterization of the time to survival and the covariates.  As a result, these models can only accommodate certain types of non-proportionality, such as non-proportional hazards that do not cross.

In the Bayesian context, non-proportionality  has been addressed in a number of ways; \cite{BerryNP} estimate piece-wise constant hazard rates for each of the treatment groups, and \cite{NB2014} extends a frequentist model by \cite{YP05} that estimates short and long term hazard ratios for crossing hazards.  \cite{Hennerfeind} have developed a survival model that incorporates functions for time-varying covariate effects into the hazard rate using Bayesian P-spline priors, while \cite{CaiandMeyer} develop a Bayesian stratified proportional hazards model using a mixture of B-splines.  Further, \cite{Iorio} use a dependent Dirichlet Process (DDP) to non-parametrically estimate non-proportional hazards.  We compare and discuss many of these models in Section \ref{sec:modelcheck}.

The MRH model is closely related to the Polya tree \citep{ferguson74, polyaLavine}. The Polya-tree prior is an infinite, recursive, dyadic partitioning of a measurable space $\Omega$. (Although in practice this process is terminated at a finite level $M$, resulting in ``finite" Polya trees.)  Polya trees have been adapted for modeling survival data in a number of ways (for example, see \cite{polyaSurvMuliere, polyaSurvHansonJohnson, polyaSurvHanson, polyaSurvZhao}).  A stratified Polya tree prior is developed by \cite{polyaStratifyZhao}, with the tree centered at the log-logistic parametric family.  The ``bins" of the Polya tree are fused together in the ``optional Polya tree" (OPT), developed by \cite{polyaOPTWong}, with fusing performed through randomized partitioning of the measurable space and a variable that allows for the stopping of the partitioning in different subregions of the tree.  The ``rubbery Polya tree" (rPT), developed by \cite{polyaRPTNB}, smooths the partitioned subsets by allowing the branching probabilities to be dependent within the same level, implemented through a latent binomial random variable.  

The MRH prior is a type of Polya tree; it uses a fixed, pre-specified partition, and controls the hazard level within each bin through a multi-resolution parameterization.  This parameterization allows parameters to differ across bins and levels of the tree in such a way that  the marginal priors at higher levels of the tree are the same, regardless of the priors at lower levels of the tree. The MRH model is capable of producing group-specific non-proportional hazards estimates (in the presence of proportional hazards covariates), allows for a data-driven fusing of bins (called ``pruning"), and includes parameters that can control the smoothness and correlation between intervals. 

%%%%%%%%%%%%%%%%%%%%%%%%%%%%%%%%%%%%%%%%%%%% %%%%%%%%%%%%%%%%%%%%%%  
\subsection{MRH Methodology for Mon-proportional Hazards (NPMRH)}
%										METHODOLOGY
%%%%%%%%%%%%%%%%%%%%%%%%%%%%%%%%%%%%%%%%%%%% %%%%%%%%%%%%%%%%%%%%%%  

The basic MRH was extended in \cite{Dukic} into  the hierarchical  multi-resolution (HMRH)  hazard model,  capable of modeling non-proportional hazard rates in different subgroups jointly with other proportional predictor effects. Like HMRH, the methodology in this paper allows for group-specific hazard functions, but adds the pruning methodology of \cite{Yprune} for individual hazard rates. The  pruning algorithm detects consecutive time intervals  where failure patterns are statistically similar, increasing estimator efficiency and reducing computing time.  The resulting method produces  computationally stable and efficient inference,  even in periods with sparse numbers of failures, as may be the case in studies with long follow-up periods \citep{Yprune}.   Details of the MRH prior and the pruning method are found in Appendix A.

\subsubsection{NPMRH Likelihood Function} 
The biochemical failure time data is subject to right-censoring: a patient's biochemical failure time is considered right-censored if it has not been observed before the end of the study period (time $t_J$), or the censoring time ($t_{cens}$, where $t_{cens} < t_J$).  We denote $T_i$ as the minimum of the observed time to biochemical failure or the right-censoring time for subject $i.$   Each patient belongs to one of the $\mathcal{L}$ covariate (for example a combination of treatments) strata, and within each stratum we employ the proportional hazards assumption such that: $$h_\ell(t \mid X, \vec\beta) = h_{base, \ell}(t)\exp\{\mathbf  X' {\vec \beta}\}.$$ 
Here, $h_\ell$ denotes the baseline hazard rate for treatment strata $\ell$,  ${\mathbf X}$ represents the $z \times n_{\ell}$  matrix of $z$ covariates (other than those used for stratification) for the $n_{\ell}$ patients in the stratum $\ell$, while ${\vec \beta}$ denotes the $z \times 1$ vector of the covariate effects. 

For subject $i$ in  stratum $\ell$ who has an observed failure time at $T_{i,\ell} \in [0, t_J)$, the likelihood contribution is: 
\begin{align*}	
	L_{i, \ell}(T_{i, \ell} \mid X_{i, \ell}, \vec \beta) = h_{base, \ell}(T_{i, \ell})\exp(X_{i,\ell}'\vec \beta)S_{base, \ell}(T_{i, \ell})^{\exp(X_{i,\ell}'\vec \beta)},
\end{align*}
where $X_{i,\ell}$ is that subject's covariate vector, and $S_{base, \ell}$ is the baseline survival function for the stratum $\ell$.
Similarly, for a subject in stratum $\ell$ whose  failure time is greater than the censoring time, the likelihood contribution is: 
\begin{align*}	
	L_{i, \ell}(T_{i, \ell} \mid X_{i, \ell}, \vec \beta) = S_{base, \ell}(T_{i, \ell})^{\exp(X_{i,\ell}'\vec \beta)}.
\end{align*}
Thus, the likelihood for all $n$ patients in all $\mathcal{L}$ strata together ($n = \sum_{\ell = 1} ^{\mathcal{L}} n_{\ell}$) is
\begin{eqnarray*}
	L(\mathbf{T \mid \vec \beta, H, R_{m,p}, X}) = \prod_{\ell = 1}^{\mathcal{L}}\prod_{i \in \mathcal{S}_\ell}
			\left[h_{base, \ell}(T_{i, \ell})e^{X_{i, \ell}'\vec \beta}\right]^{\delta_{i, \ell}}S_{base, \ell}(T_{i, \ell})^{e^{X_{i, \ell}'\vec \beta}},
\end{eqnarray*}
where ${\mathcal{S}_\ell}$ denotes the set of indices for subjects belonging to the stratum $\ell$,  and $\delta_{i,\ell} =  1$ if subject $i$ had observed biochemical failure, and 0 if censored.
In this model, the $\mathcal{L}$ hazard rates are estimated jointly with all the covariate effects. The estimation algorithm is  performed two steps: the pruning step and the Gibbs sampler routine.  Details are provided in Appendix B.

\subsubsection{Inference for Non-proportional Effects}

A non-proportional covariate effect can be described as the log of the hazard ratio between different covariate strata in each bin. For simplicity, let us assume that the model has only one non-proportional effect predictor with $\ell$ categories -- for example,  $\ell$ treatment groups. Let $\alpha_{\ell, \ell+1}(t)$ denote the hazard  effect of treatment group $\ell+1$ with respect to treatment group $\ell$. Then  $\alpha_{\ell, \ell+1}(t)$ can be thought of as a time-varying effect, which  is constant within each time bin, but changes across time bins. In other words, 
\begin{align*}
	\alpha_{\ell, \ell+1}(t) & = \big[\alpha_{\ell, \ell+1}(t_1-t_0), \dots, \alpha_{\ell, \ell+1}(t_J-t_{J-1})\big]'\\
		& = \log\left(\big[d_{1,\ell+1}/d_{1,\ell}, \dots, d_{J,\ell+1}/d_{J,\ell}\big]'\right),
\end{align*}
where $d_{j,\ell}, j = 1, \dots, J, \ell = 1, \dots, \mathcal{L}$ represents the hazard increment in the $j^{th}$ bin of the $\ell^{th}$ strata, and is defined as $d_j = \int_{t_{j-1}}^{t_j} h(s) ds = H(t_j) - H(t_{j-1})$ (see Appendix A for details). For the biochemical failure model, the time-varying effect of treatment is thus:
\begin{align*}
	\alpha_{0, 1}(t) = \log\left(\big[d_{1,1}/d_{1,0}, \dots, d_{J,1}/d_{J,0}\big]'\right),
\end{align*}
where 0 represents the short-term androgen deprivation therapy group, and 1 represents the long-term androgen deprivation therapy group.  The marginal posterior distribution of this time-varying effect of treatment can therefore be obtained directly from the joint posterior distribution for the hazard increments $d$.

%%%%%%%%%%%%%%%%%%%%%%%%%%%%%%%%%%%%%%%%%%%% %%%%%%%%%%%%%%%%%%%%%%  
%											
% RESULTS
%%%%%%%%%%%%%%%%%%%%%%%%%%%%%%%%%%%%%%%%%%%% %%%%%%%%%%%%%%%%%%%%%%  

\section{Analysis of RTOG Prostate Cancer Clinical Trial Data}

The main goal of our analyses of the time biochemical failure was to infer how the effects of AD therapy impacted the failure time, and if and how they may have changed over the course of treatment and the subsequent  10 year  post-diagnosis follow-up. Of particular interest was to assess whether the benefit of longer over shorter duration androgen deprivation (AD) therapy was persistent over time or if it diminished in the late follow-up. Additionally of interest was whether the biochemical failure hazard rate for  the long-duration AD (+24m) group increased later in  follow-up, indicating that failures were deferred in time rather than avoided. Inference in the later time periods is more challenging however, as the RTOG 92-02 clinical trial exhibits sparsity of events towards the end of the follow-up time as is typical of long-term studies. Fewer observed biochemical failures occurred in the later periods, with only 13\% of the subjects having observed biochemical failure after 4.9 years (the median time to biochemical failure), and only 1.5\%  after 10 years.  

Several previous studies have  used time aggregated summaries (i.e., survival curves, cumulative incidence) to  estimate cumulative biochemical failure risk over time \citep{NatHist1, NatHist2}. However, previous analyses specifically examining the hazard rate of biochemical failure are relatively scarce.  These analyses have been limited to the clinical literature and used intuitive summaries to approximate the annual hazard, for example, by calculating the annual number of events divided by the number at risk \citep{HRAmling, HRDillioglugil, HRHanlon, HRWalz}. These analyses provide basic, useful information on the patterns of hazard of biochemical failure over time, and have helped identify higher and lower periods of hazard for specific patient groups.  However, the straightforward methods used do not provide smoothed estimates of the hazard rate over time, and the joint estimation of the effects of multiple covariates on hazards was not considered.  In addition, during periods of time when no biochemical failures were observed, these estimation procedures calculated the hazard rate to be zero.  

Thus, the analyses here  investigate the effects of treatment, age, Gleason scores, PSA levels, and T-stage  at diagnosis on the time to biochemical failure, allowing for possible non-proportional treatment effects.  In order to provide a thorough investigation of the treatment hazard ratio over time and to determine the effects of  different modeling and smoothness assumptions on the estimate, we present a suite of models ranging from simple parametric models to more complex non-parametric models.  In addition to the standard and pruned MRH models, we also employ a parametric non-proportional hazards Weibull model, two piece-wise exponential models \citep{zelen,Ibrahim}, an extended Cox model that allows for time-varying covariate effects \citep{timereg}, a Dependent Dirichlet Process survival model \citep{Iorio}, and a semi-parametric Bayesian accelerated failure time model \citep{bayesSurv}.  All MRH models were implemented using the ``MRH'' R package \citep{MRHR}. 

The following covariates were examined in the models:
\begin{itemize}
	\item Treatment (+0m vs +24m AD therapy) \\[-4ex]
	\item Age (categorized: less than 60 years, 60 to 70- years, 70 to 80- years, and 80 years or older) \\[-4ex]
	\item Gleason scores (categorized: low grade, intermediate grade, and high grade, corresponding to scores between 2-4, 5-7, and 8-10, respectively) \\[-4ex]
	\item PSA levels (log transformed and then centered at the mean log value equal to 3) \\[-4ex]
	\item T-stage (binary: stage 2, or stage 3/4)
\end{itemize} 
The baseline (reference) group comprises subjects who received  no additional androgen deprivation therapy (+0m), had an intermediate Gleason score, were below age 60 at study entry, and had a T-stage equal to 2. 

\subsection{MRH Results}
\label{section:MRHres}

The time resolution $M = 6$  was chosen for this analysis in order to provide a fine grain examination of biochemical failure patterns over the course of more than 13 years. The resulting  $J = 64$  time intervals, partitioning the time axis into bins of length 2.5 months, allowed us to investigate detail in the biochemical failure hazard rate that is useful to clinical practice.  The full 64-bin MRH model with non-proportional treatment hazards (the ``NPMRH-0" model) was compared to two pruned MRH models with non-proportional treatment hazard rates, one with all 6 levels of the MRH tree pruned (``NPMRH-6"), and one with only the bottom 3 levels of the MRH tree pruned (``NPMRH-3").  Pruning the 64-bin model allowed us to fuse bins where the failure rates were statistically similar (reducing the number of model parameters), which in turn helped us identify periods of time where the hazard rates were flat and where treatment effects remained steady.  To test whether the non-proportionality of hazards  was indeed warranted in the model,  we also examined a pruned MRH model (a 6-level model with the bottom 3 levels subject to pruning) with the treatment effect included under the proportional hazards assumption (PHMRH).  

Five separate Markov chain Monte Carlo (MCMC) chains were run for each model, each with the burn-in of 50,000, leaving a total of 150,000 thinned iterations in each chain for analysis.  Convergence was determined through the Geweke diagnostic \citep{geweke}, graphical diagnostics, and Gelman-Rubin tests \citep{gelmanrubin, gelmanrubinadd}. These diagnostics  are automatically presented when using the MRH R package \citep{MRHR}.  Point estimates for the MRH models were calculated as the median of the marginal posterior distribution of each parameter. Central credible intervals  were used for inference. 

\begin{figure}[htbp] %%%% MRH estimates, smoothed
	\centering
	\includegraphics[width=5.5in]{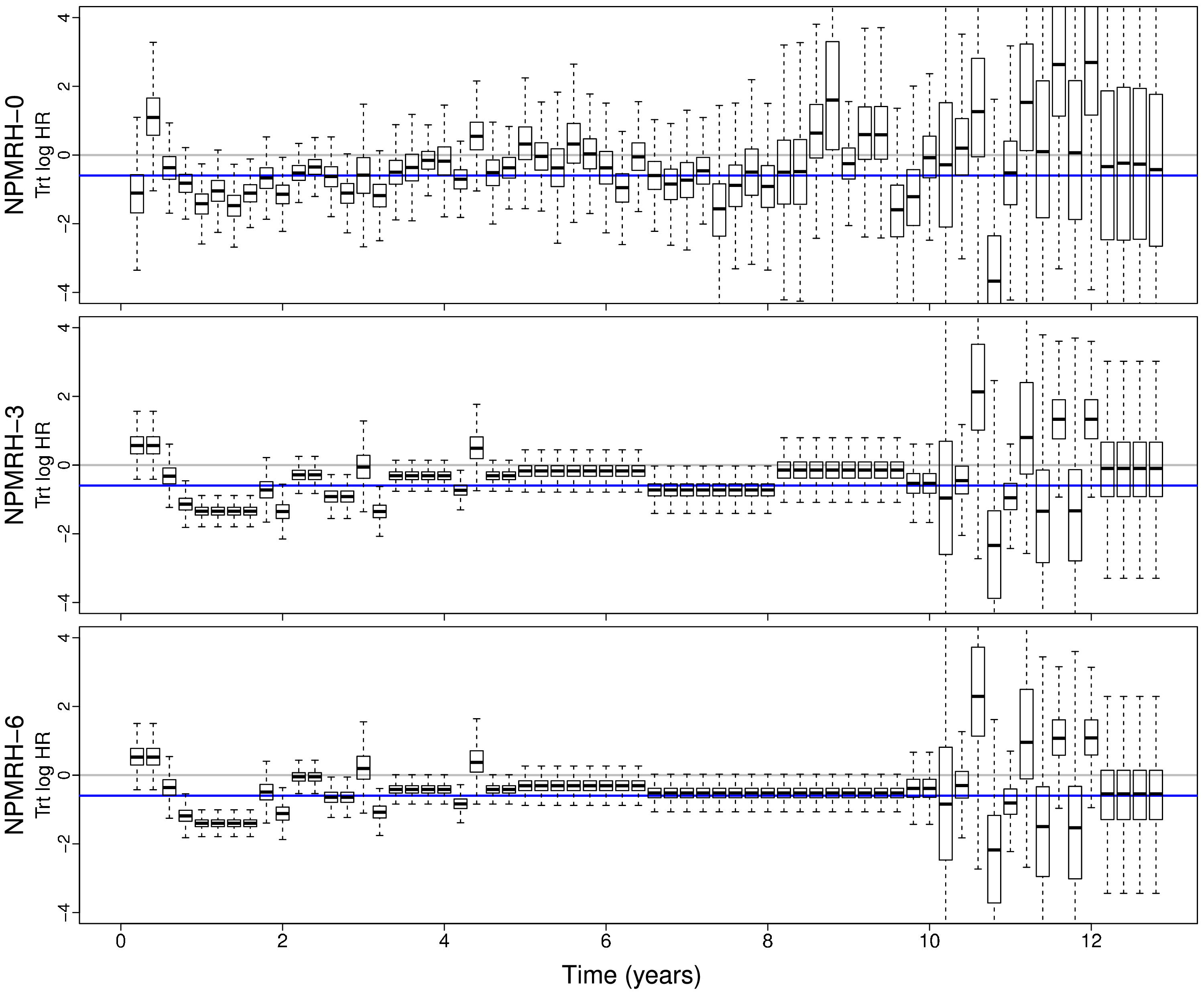}
	\includegraphics[width=3in]{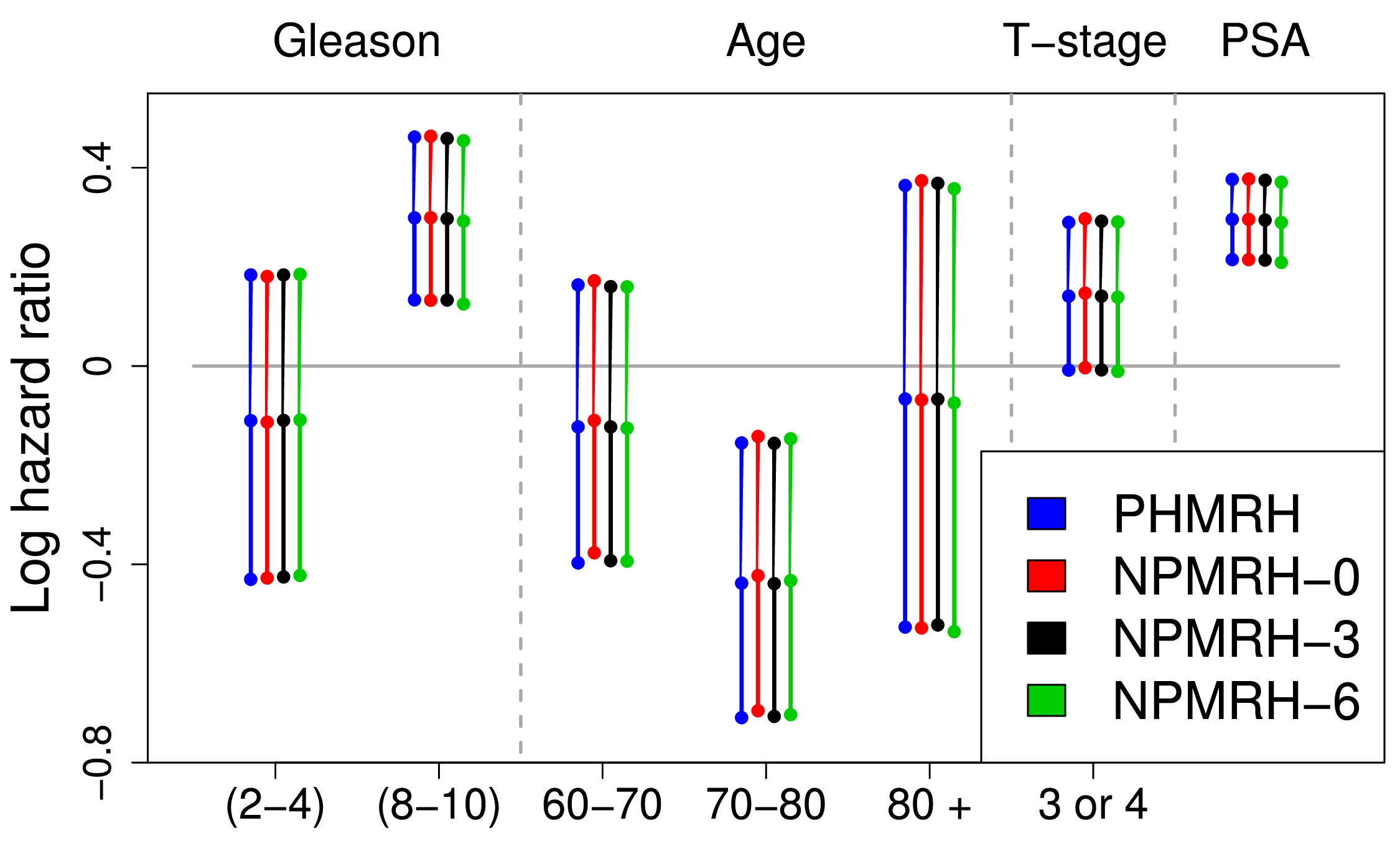}
	\includegraphics[width=3in]{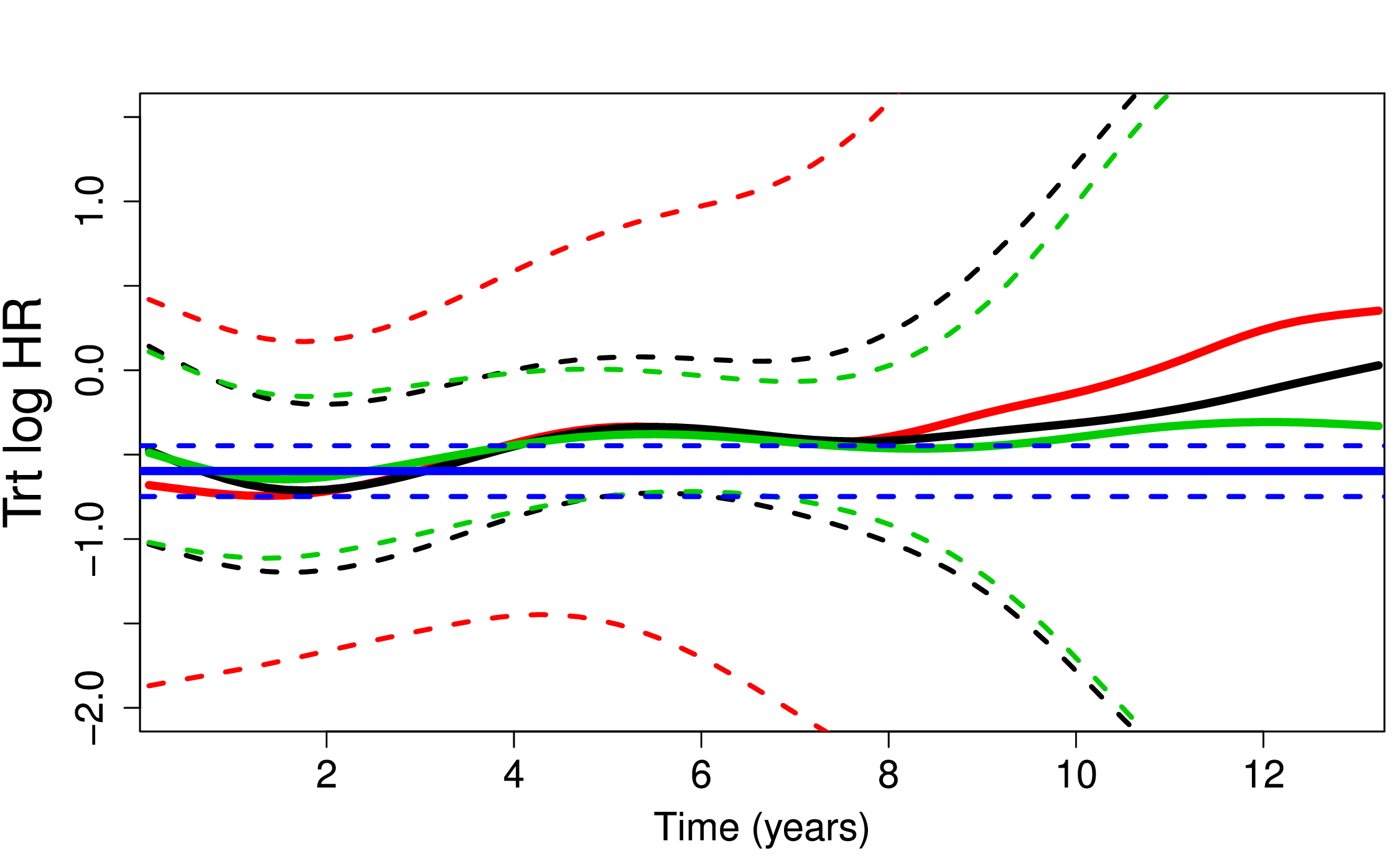}
	\caption{\footnotesize Comparison of parameter estimates produced by the different  MRH models.  
TOP: Caterpillar plots of the treatment log-hazard ratio  for the three non-proportional MRH models with different levels of pruning.  The model on the top has no pruning (NPMRH-0), and shows the most variation between and within bins.  As the level of pruning increases (NPMRH-3 and NPMRH-6), the uncertainty of the estimates decreases.  In addition, the pruned models communicate the patterns in the log-hazard ratio more clearly:  larger treatment differences are visible at the beginning and end of the study, with long periods of stability in the middle. All caterpillar plots have two reference horizontal lines: the grey line crosses the y-axis at 0, and the blue line  at -0.597 (the estimate of the treatment effects under the proportional hazards using the the PHMRH model).
BOTTOM LEFT: Covariate effect estimates and corresponding credible bounds for the MRH and NPMRH models.  The figure shows that the estimated covariate effects are very similar among all models, regardless of the level of pruning or the treatment effect proportionality assumption.  
BOTTOM RIGHT: The smoothed log  hazard ratio of the long-duration AD (+24m) group to the short-duration (+0m) group for  NPMRH-0, NPMRH-3, and NPMRH-6 models, contrasted against the estimate of treatment effect under the proportional hazards assumption (obtained from PHMRH).  The solid lines represent the log-hazard ratio estimate, and the dashed lines represent the smoothed point-wise 95\% credible intervals. The hazard rate estimates are very similar among the three models.  However, the two models that have pruned trees have narrower credible intervals, as they have a smaller number of estimated parameters.  Note that in all models, the credible bands  become large towards the end of the study, due to the decreasing frequency of observed biochemical failures as time progresses.  However, the pruned models show lower variability towards the end of the study period, as the per-bin observed failure count is higher in those models.}
	\label{fig:comparePrune}
\end{figure}

\begin{figure}[htbp] %%%% MRH estimates, smoothed
	\centering
	\includegraphics[width=6in]{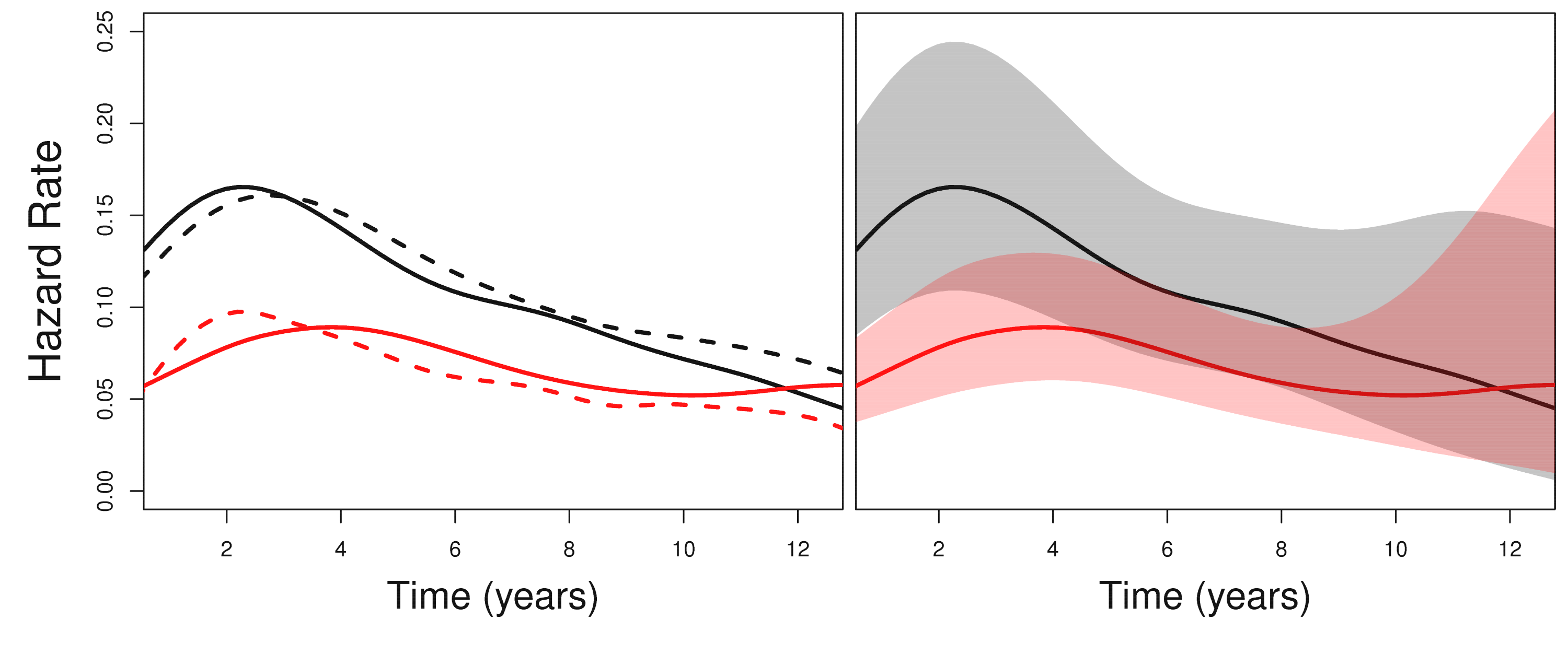}
	\includegraphics[width=6in, height=2.2in]{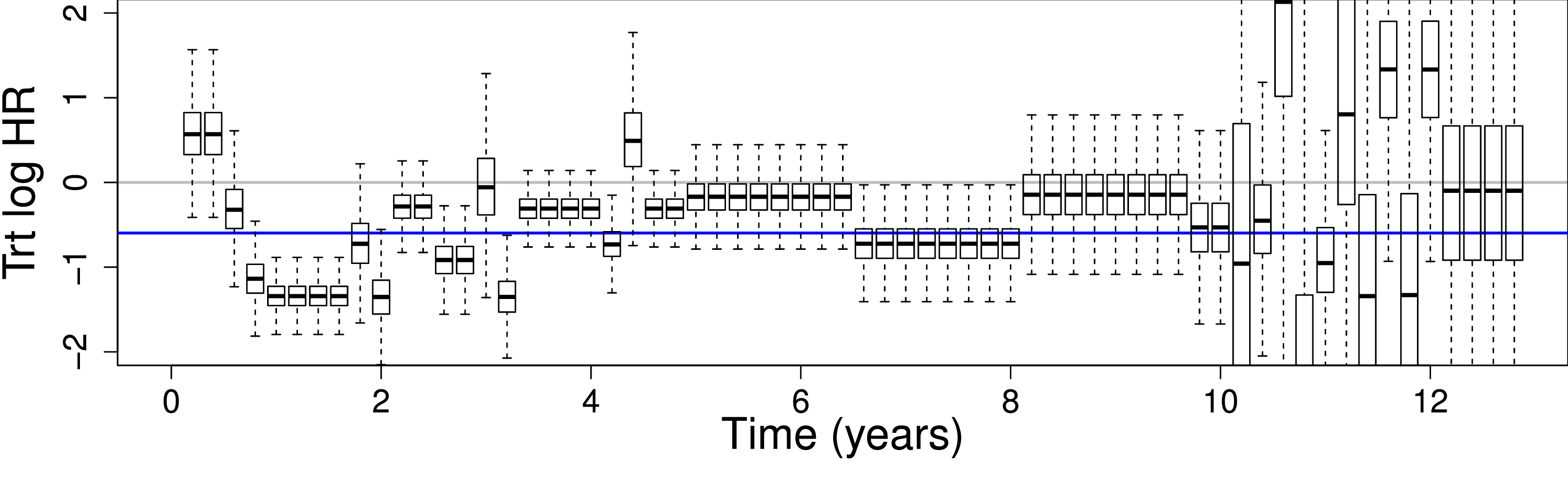}
	\caption{\footnotesize TOP LEFT: Smoothed estimated hazard rates (baseline subjects only) for the +24m AD therapy group (red) compared to the +0m AD therapy group (black).  The hazard rates estimated under the non-proportional assumption are represented with solid lines, and the hazard rates estimated under the proportional assumption are represented with dashed lines (calculating the +24m hazard rate at time $t$ as $h_0(t)\exp\{\beta_{tx}\}$).  While the estimated hazard rate for the +0m AD therapy group is similar under both the proportional and non-proportional modeling assumptions, the +24m hazard rate estimates have larger departures, with a flatter 2-year peak for the estimate from the non-proportional hazards model.  TOP RIGHT:  Smoothed estimated hazard rates (baseline subjects only) and 95\% credible interval bounds for the +24m AD therapy group (red) compared to the +0m AD therapy group (black).  The intervals are slightly narrower for the +24m treatment group when compared to the +0m treatment group, although the credible intervals for the +24m estimated hazard rate become wider at the tail end of the end of the study where few failures are observed.  BOTTOM: A caterpillar plot of the effects of long-term vs short-term treatment over time.  The grey line lies on the y-axis at 0, and the blue line lies on the y-axis at -0.597, which is the estimate of the treatment effect under the proportional hazards setting (estimate from the PHMRH model).  It is particularly apparent at the beginning of the study that the proportional hazard rate estimate for treatment is not contained in the boxplot bounds.  In addition, we can see that the boxplot medians have a lot of variation, and even change from negative to positive multiple times throughout the course of the study.  All estimates shown are from the NPMRH-3 model.}
	\label{fig:MRHhazards}
\end{figure}

One of the most notable features of our results are in the estimated log-ratio of the treatment effect, particularly in the pruned models.  In examining Figure \ref{fig:comparePrune}, we note the  interval-specific differences displayed in the caterpillar plots (top 3 plots in Figure \ref{fig:comparePrune}).  These plots reveal several important pieces of information about the treatment effects, including time periods where the treatment effects were: 1) proportional (constant) or non-proportional (changing), 2) statistically significantly different from previous periods, and 3) statistically significantly different from zero or from the proportional hazards model estimate of the treatment effect.  For example, in the NPMRH-3 model, the treatment effects remained steady between (approximately): 6 ms-2 years, 3.5-4 years, 5-6.5 years, 6.5-8 years, and 8-10 years.  However, these periods of constant estimated treatment effects were different from one another, suggesting that while the benefits of treatment lasted for a certain number of years, the degree of improvement changed (and generally declined) over the course of the study.  Between 6 months and 2 years, long-term AD therapy had an estimated 75\% improvement over short term AD therapy.  In examining the 95\% bounds of the boxplots, this estimated log-hazard ratio is statistically different than the log-ratio estimate from the proportional hazards model of $\widehat{\beta_{tx}} = -0.597$ (which translates to 45\% improvement for the +24m group).  Additionally, the estimated log-ratio in this time period is statistically significantly different than the estimated treatment effects between 5-6.5 and 8-10 years, which only showed an estimated 26\% improvement for subjects on long-term therapy (see Figure \ref{fig:MRHhazards}, bottom).  In both pruned models, the treatment effect held steady for a certain number of years, then diminished slightly, and held steady for another number of years, before diminishing in effectiveness again.  Overall, long-term AD therapy did better  in prolonging time to biochemical failure throughout most of the first 10 years of the study, despite the fact that in certain periods the log-ratio is not statistically significantly different from zero.  

The results of all the MRH models provide two important insights: 1) The proportional hazards assumption indeed did not hold for treatment effects (agreeing with the Cox model test), and there were in fact periods of time where the estimated effects are statistically significantly different from each other, and 2) On average, the subjects on +24m of AD therapy experienced benefits for at least 10 years post treatment.  

In Figure \ref{fig:comparePrune} (bottom right), we present the smoothed version of the caterpillar plots above, illustrating the overlap of the credible regions around the estimated log-hazard ratio for the four different models. The smoothing was done using a cubic smoothing spline (with 5 degrees of freedom, 53 knots, and a smoothing parameter equal to 0.82), which was implemented via the \texttt{smooth.spline()} function in R. While the caterpillar plots are useful for identifying specific interval differences in the treatment effect, these smoothed plots emphasize the difference in the uncertainty among the models and the different shapes of both the estimated effects and their credible intervals.  For example, we see that among the NPMRH models, the unpruned model (NPMRH-0) has the widest credible interval bands, while the fully pruned model (NPMRH-6) has the narrowest credible interval bands, which is due to the smaller number of estimated parameters and larger failure counts per bin in the pruned model.  While the PHMRH model clearly has the narrowest credible region, the constant parameter estimate cannot identify periods of increased or decreased long-term treatment benefit.  This discrepancy is particularly visible in the last third of the study, where the benefits of long-term treatment seem to be decreasing. 

The estimates and their 95\% credible intervals for the time-invariant effects  (effects of age, Gleason scores, PSA measures, and T-stage) are almost identical among all the MRH models, and are shown in the ``cat-scratch'' plot in Figure \ref{fig:comparePrune}, bottom left.  

In all models, estimates of the biochemical failure hazard rate for each treatment group showed an increase in the first two to four years, with a steady decline afterwards (see Figure \ref{fig:MRHhazards}, upper left).  However, subjects who received 24 months of additional AD therapy had a lower hazard rate than those who did not, with a flatter peak between 2 and 4 years.  The non-proportionality between the hazards is particularly visible when compared to the results from the proportional hazards model.  While both the NPMRH-3 and PHMRH models show similar estimated hazard rates for the +0m AD therapy group, the estimated hazard rates for the +24m group had significant departures in the first four to five years of the study, as well as in the last two years of the study.  It does appear that, while long-term treatment effects diminished over time, biochemical failure was not simply postponed for the +24m group, but the risk was in fact reduced even over a longer period of time.

Time-invariant effect estimates show that an increase in Gleason scores was associated with an increased hazard rate, with a statistically significant difference between baseline subjects and subjects with scores greater than 8  (HR = 1.35, 95\% CI: 1.14, 1.59).  The hazard of biochemical failure decreased with age, although significant differences were only observed for subjects between 70 and 80 years old and baseline subjects (HR = 0.64, 95\% CI: 0.49, 0.86).  As expected, subjects with a T-stage of 3 or 4 had a higher  hazard of biochemical failure compared to subjects with a T-stage equal to 2 (HR = 1.15 , 95\% CI: 0.99, 1.34). Similarly, for every point increase in PSA scores on the log scale (a 2.7 factor increase in PSA measures on the standard PSA scale), there was a statistically significant 34\% increase in the hazard rate. (See Table \ref{tab:mrhres}, Figure \ref{fig:comparePrune} bottom left.) 
\begin{table} %%% Covariate results table for MRH models %%%
	\center
	\begin{tabular}{|ll|c|c|c|}
%model1
		\hline 
		&&&&\\
		\multicolumn{2}{|c|}{Coefficient}&$\widehat{\beta}$&$\widehat{HR}$&95\% CI for $\beta$\\\hline\hline
		\multirow{2}{*}{Gleason score}&2-4&-0.11&0.90&(-0.43, 0.18)\\
		&8-10&0.30&1.35&(0.13, 0.46)\\
		\hline
		\multirow{2}{*}{Age}&$60-70^-$&-0.12&0.88&(-0.39, 0.16)\\
		&$70-80^-$&-0.44&0.64&(-0.71, -0.16)\\
		&80 or older&-0.07&0.94&(-0.52, 0.37)\\
		\hline
		\multicolumn{2}{|l|}{T-stage 3 or 4}&0.14&1.15&(-0.01, 0.29)\\
		\hline
		\multicolumn{2}{|l|}{log(PSA), centered}&0.29&1.34&(0.21, 0.37)\\
		\hline
	\end{tabular}
	\caption{\footnotesize Estimated covariate effects for the NPMRH-3 model.  Baseline subjects are those who have a Gleason score between 5 and 7, are less than 60 years of age, and have a T-stage equal to 2.  The $\widehat{\beta}$ column contains the estimates of the log of the hazard ratios, and the $\widehat{HR}$ column contains the coresponding estimated hazard ratios.  Model estimates for all the  different MRH (with and without pruning) are very similar, as can also be observed in Figure \ref{fig:comparePrune}.}
	\label{tab:mrhres} 
\end{table}

\begin{figure} %%%% Survival densities
	\centering
	\includegraphics[width=7in]{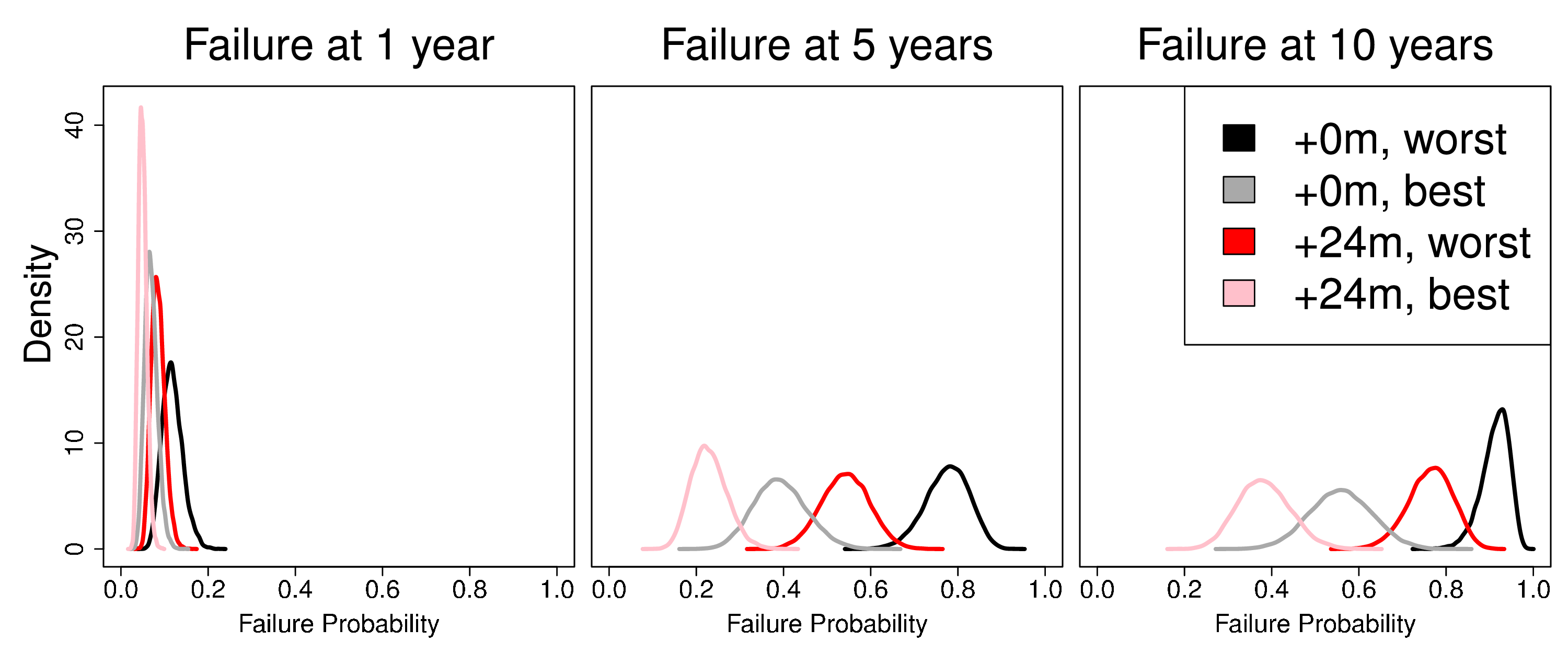}
	\caption{\footnotesize Smoothed posterior predictive densities of biochemical failure at 1, 5 and 10 years post-diagnosis, stratified by treatment type and hypothetical patient covariate profile. A subject with a ``worst" profile had a Gleason score $\ge 8$, a T-stage equal to 3 or 4, and a PSA score equal to 1 standard deviation greater than the mean (PSA $\approx 52$).  A subject with a ``best" profile had a Gleason score $\le 4$, a T-stage equal to 2, and a PSA score equal to 1 standard deviation below the mean (PSA $\approx 8$).  At one year post-diagnosis, the predictive densities of biochemical failure were very similar for all groups.  However, at 5 and 10 years the densities became more spread out.  A worst profile subject on +0m of AD therapy had the highest predictive probability of biochemical failure, while a best profile subject on +24m AD therapy had the lowest predictive probability of biochemical failure.  While all densities overlapped at one year, at 5 and 10 years very little overlap remained between the best and worst profile within the same treatment regimen.  In addition, at all three time points, the predictive probability of biochemical failure was higher for the worst profile subjects (regardless of treatment) than the best profile subjects.  Smoothed density estimates were calculated using \texttt{density()} in R.}
	\label{fig:survCurveComparison}
\end{figure}

The probability of biochemical failure at 1, 5, and 10 years can be observed in Figure \ref{fig:survCurveComparison}, which shows the smooth  posterior predictive probability densities of biochemical failure, stratified by treatment type for hypothetical subjects with a``worst" or ``best" covariate profile.  A subject with a ``worst" profile had a Gleason score $\ge 8$, a T-stage 3 or 4 tumor, and a PSA score equal to 1 standard deviation greater than the mean (PSA $\approx 52$).  A subject with a ``best" profile had a Gleason score $\le 4$, a T-stage equal to 2, and a PSA score equal to 1 standard deviation below the mean (PSA $\approx 8$).  At one year post diagnosis, we see that the posterior predictive densities were very similar among the four groups, all concentrated between 0 and 20\%.  However, by the 5-year post-diagnosis mark, the failure densities were very different.  A worst profile subject on +0m AD therapy had failure probability centering around 80\%, while a best profile subject on +24m of AD therapy had failure probability centering around 20\%.  It can also be observed that a worst profile subject had higher failure probability than a best profile subject, regardless of treatment type.  The failure probability at 10 years post-diagnosis is perhaps the most telling, with a worst profile subject on +0m AD therapy having a failure probability ranging from 80-100\%, which is a narrower range when compared to the other groups.  Meanwhile, a best profile subject on +24m therapy had failure rates centering around 40\%, over a wider interval from approximately 20\% to 60\%.  While all posterior predictive densities overlapped at one year, at 5 and 10 years there was only a small amount of overlap between the best and worst profile subjects within the same treatment regimen.

\subsection{Model Checking and Comparison}\label{sec:modelcheck}
To assess the impact of different modeling and smoothness assumptions on the hazard of time to biochemical failure, we compared the four MRH models to each other as well as to other models, including the Cox proportional hazards model, a parametric non-proportional hazards Weibull model, two piece-wise exponential (PE) models, a dependent Dirichlet Process (DDP) survival model, and a semi-parametric Bayesian accelerated failure time (AFT) model, allowing for time-varying treatment effects in all models.  In addition, we performed a sensitivity analysis to the parameter $k,$ which controls the smoothness in the MRH tree prior (\cite{Bouman}, see Appendix A for details).  When applicable, models were compared through a goodness-of-fit measure (defined in Section \ref{sec:gofsec}), as well as via information criteria including  BIC \citep{refBIC}, AIC \citep{refAIC}, and DIC \citep{refDIC, refDIC1}.  %For the MRH models, the effective number of parameters used in the calculation of the BIC and AIC values was equal to $10+2^M-1$, with 1 for $H$ (the cumulative hazard), 2 for $a$ and $\lambda$ (in the priors for $H$), $2^M-1$ for the $R_{m,p}$ parameters, and 7 for the number of fixed covariates in the model.  The $H, a, \lambda,$ and $R_{m,p}$ counts were multiplied by two in the NPH models, and if any $R_{m,p}$ parameters were pruned, they were deducted from the total count.

%%%%%%%%%%%%%% Model comparison
\subsubsection*{Cox Proportional Hazards Model}

\begin{SCfigure}% Schoenfeld residuals
	\includegraphics[width=3in]{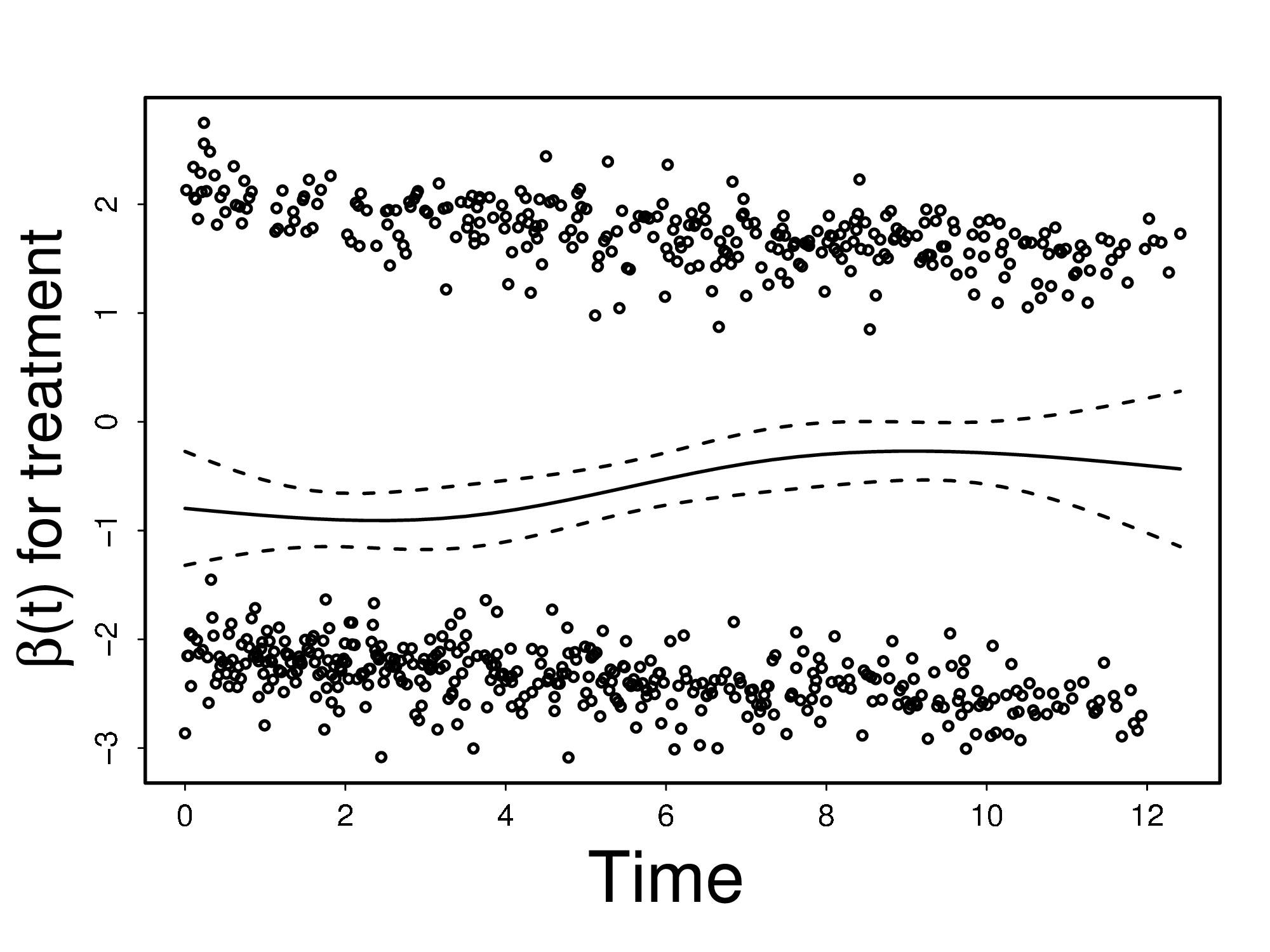}
	\caption{\footnotesize  Schoenfeld residuals for the treatment effect, with a weighted least squares line. Deviation from linearity indicates that the treatment effect does not fit the proportional hazards assumption. \vspace{.5in}}	
\label{fig:schoenlog}
\end{SCfigure}

Because the Cox proportional hazards model is widely used in the analysis of survival data, we included this model as a comparison to other non-proportional hazards model for contrast \citep{CoxPH}.  While we modeled the treatment effect under the proportional hazards assumption, it is important to note that the Schoenfeld residuals and methods presented by \cite{SchoenRes} showed evidence that the treatment effect (long-term versus short-term therapy) was not proportional over the entire study period (see Figure~\ref{fig:schoenlog}).  No other covariate effects showed evidence of non-proportionality over time.  

\subsubsection*{Cox Model Extension}
In addition to the traditional Cox proportional hazards model, we included an extended proportional hazards model with a time-varying treatment effect \citep{timereg}. This extended Cox model has a hazard rate with the form
\begin{eqnarray}\lambda(t) = Y(t)\lambda_0(t)\exp\{X^T(t)\beta(t)+Z^T(t)\gamma\},\label{eqn:timecoxHazrate_generic}\end{eqnarray}
where $(X(t), Z(t))$ is a $(p+q)$-dimensional covariate, $\beta(t) = (\beta_1(t), ..., \beta_p(t))$ is a $p-$dimensional time-varying (i.e. NPH) regression coefficient that is estimated non-parametrically, and $\gamma$ is the q-dimensional regression parameter for the PH covariate effects.  An implementation of this model was performed using the \texttt{timecox()} function in the ``timereg" package \citep{timeregPkg}, where parameters are estimated using score equations, and the optimal smoothing parameter was chosen based on the lowest -2*log-likelihood value and the lowest GOF scores.  

%$\beta(t)$ estimates are reported as cumulative estimates over the study period:
%\begin{eqnarray} B(t) = \int_0^t \beta(s) ds.\label{eqn:timecoxCumul}\end{eqnarray}
%The smoothed, de-cumulated estimates for $\hat{\beta}(t)$ can be calculated using functions provided in the ``timereg" package.  The }

\subsubsection*{Accelerated Failure Time Model}

Accelerated failure time models are an alternative way to investigate the effects of non-proportional hazards.  In our analysis, we used a Bayesian AFT model \citep{bayesSurv}.  This model can accommodate more complex clustered, interval-censored survival data, with the log of the survival  times is modeled as:
$$\log(T_{i,l}) = \beta'x_{i,l}+b_i'z_{i,l}+\epsilon_{i,l}, i = 1, \dots, N, l = 1, \dots, n_i,$$
where $T_{i,l}$ is the event time of the $l^{th}$ observation of the $i^{th}$ cluster. The model estimates the effects, $\beta = (\beta_1, \dots, \beta_p)$, of the fixed effects $x_{i,l}$ and the i.i.d random effects $b_i = (b_{i,1}, \dots, b_{i,q})^T$.  The fixed and random covariate effects are modeled using the classical Bayesian linear mixed model approach (such as \cite{BayesianMixedMods}), and the hazard rate is approximated by normal mixtures.  For the purposes of the prostate data, we omit any clustering effects.

\subsubsection*{Dependent Dirichlet Process Survival Model}

As the most flexible alternative, we also consider a non-parametric Bayesian model that can accommodate non-proportional hazards.  This model is based on the ANOVA Dependent Dirichlet Process (DDP) model presented in \cite{IorioAnova}, that has been extended to the survival analysis setting \citep{Iorio}.  The DDP survival model performs survival regression based on a Dirichlet Process prior.  

The set of the random probability distributions or functions are dependent in an ANOVA-type fashion: If $\{F_x, x \in X\}$ is the set of random distributions indexed by the categorical covariates $x = (x_1,\dots,x_p)$, and the collection of random distributions is defined as
$$F_x = \sum_{h = 1}^\infty p_h\delta(\theta_{xh}),\text{ for each }x \in X,$$
with $\sum_{h = 1}^\infty p_h = 1,$ and $\delta(y)$ representing a point mass at $y,$ then dependence is introduced by modeling the locations $\theta_{xh}$ through the covariates (as explained in \cite{IorioAnova} and \cite{Iorio}).  This model allows all group-specific hazards to be modeled non-proportionally, and covariate effects are interpreted in the standard ANOVA manner.  The model is also capable of accommodating continuous covariates.

Because of the greater flexibility with the DDP survival model, this model is often unable to estimate all desired covariate effects.  As a result, we present this model on a reduced set of variables that were found to be significant in other models: treatment, a high Gleason score, age between 70 and 80 years, and log(PSA).  

\subsubsection*{Non-Proportional Hazards Weibull Model} 
The non-proportional effects Weibull model was designed with separate Weibull hazard rates for each treatment group and proportional hazard covariate effects shared among both treatment groups.  Parameter estimates for this model were obtained using numerical  optimization of the likelihood function:
\begin{align*}
	L & = \prod_{\ell = 1}^2\left[ \prod_{i \in tx_\ell} \left(\kappa_\ell \lambda_\ell \left(\lambda_\ell T_{i, \ell}\right)^{\kappa_\ell-1}\exp\left\{X_{i, \ell}'\vec\beta\right\} \right)^{\delta_{i, \ell}}
			\exp\left\{-(\lambda_\ell T_{i, \ell})^{\kappa_\ell}\exp\left\{X_{i, \ell}'\vec\beta\right\}\right\}\right],
\end{align*}
where $\vec\beta$ are the covariate effects modeled under the proportional hazards assumption.  The estimate of the log-hazard ratio of the non-proportional effect of treatment at time $t > 0$ in the Weibull model was then obtained as
\begin{eqnarray*}
	\alpha_{W}(t) = \log\left(\frac{\kappa_{1}\lambda_{1}\left(\lambda_{1} t\right)^{\kappa_{1}-1}}
				{\kappa_{0}\lambda_{0}\left(\lambda_{0} t\right)^{\kappa_{0}-1}}\right),
\end{eqnarray*}
where group 0 is the short-term treatment group, and group 1 is the long-term treatment group.  The non-proportional hazards Weibull model parameter estimation was not performed using any available software packages, but is available on request from the authors. 

%The effective number of parameters  used in the BIC and AIC calculations was equal to 11, with 2 each for the $\kappa_\ell$ and $\lambda_\ell$ ($\ell = 1,2$) and 7 for the number of fixed covariates in the model.

\subsubsection*{Piece-wise Exponential Models}
The piece-wise exponential (PE) model is a commonly used frequentist semi-parametric model for joint estimation of the hazard rate and covariate effects (for example, see \cite{PEexample}).  It is similar to the MRH model in that both assume constant hazard rates within a time bin $j$ ($j = 1,\dots, J$), but it does not have the multi-resolution aspects of MRH.  

As with the non-proportional hazards Weibull model, we fit a PE model with  separate hazard rates for each treatment group, and shared proportional hazards effects among all subjects. 
If we let  $\lambda_{j,\ell}$ represent the constant hazard rate in the $j^{th}$ bin for the $\ell^{th}$ treatment group ($j = 1,\dots,J,$ and  $\ell = 1,2$), then the piece-wise exponential likelihood can be written as:
\begin{align*}
	L(\mathbf{T \mid \vec \beta, X, \delta, \lambda}) &  = \prod_{i=1}^n\prod_{\ell = 1}^2 \prod_{j = 1}^J \left(\lambda_{j,\ell} \exp\{X_{i}'\vec\beta\}\right)^{\delta_{ij,\ell}} 
						\exp\left\{-\omega_{ij,\ell} \lambda_{j,\ell}\exp\{X_{i}'\vec\beta\}\right\},
\end{align*}
where
\[ \delta_{ij,\ell} = \left\{ \begin{array}{ll}
         1 & \mbox{if subject $i$ is in treatment group $\ell$ and has a failure in time bin $j$}\\
        0 & \mbox{otherwise},\end{array} \right. \] 
 \[ \omega_{ij,\ell} = \left\{ \begin{array}{ll}
         t_{j,\ell}-t_{j-1,\ell} & \mbox{if subject $i$ is in treatment group $\ell$ and } T_{i,\ell} > t_j,\ell\\
          Ti-t_{j-1,\ell} & \mbox{if subject $i$ is in treatment group $\ell$ and } T_{i,\ell} \in [t_{j-1,\ell}, t_{j,\ell}]\\
        0 & \mbox{otherwise}.\end{array} \right. \] 
      
To make the PE model comparable with the pruned  MRH models, we use a data-driven method to select the optimal number of bins $J,$ as well as the optimal bin width(s).  In addition,  we modify the standard PE approach slightly in order to overcome a common obstacle in the estimation of the variance. Namely, given that the Fisher Information for the hazard rate in bin $j$ for group $\ell$ can be derived  as:
$$ I(\lambda_{j,\ell}) = -\frac{\sum_{i=1}^n \delta_{ij,\ell}}{\lambda_{j,\ell}^2},$$
bins with no observed failures  will yield $ I(\lambda_{j,\ell})$ of zero,  making the Fisher Information matrix singular.  In such instances,  we have remedied this issue by (repeated) merging of the bins with no observed failures into the adjacent bins to the left. With that modification, for each of the hazard rates, we find the PE model with the optimal number of bins and bin widths based on an information criterion such as AIC \citep{refAIC}, in two ways: 

\begin{enumerate}
	\item Equal-bin model:  The ``equal-bin"  PE model partitions the time axis evenly into $j$ bins  (where $j = 2,\dots,J$). Among the $J-1$ equal-bin PE models, we retain the model that has the lowest AIC value.  
	\item Quantile-bin model:  The ``quantile-bin" PE model partitions the time axis  into $j$ quantiles ($j = 2,\dots,J$). Among the $J-1$ quantile-bin PE models, we retain the model that has the lowest AIC value.  

\end{enumerate}

In the RTOG 92-02 data set, the final equal-bin PE model had 17 bins for the +0m treatment group hazard rate and 18 bins for the +24m treatment group hazard rate.  The last three bins were combined for the +0m treatment group, and the last two bins had to be combined  for the +24m treatment group hazard rate.  (Note that as the result, not all bins were of equal length due to the combined bins at the end of the study).  The final quantile-bin model had 24 bins  for the +0m treatment group hazard rate and 25 bins for the +24m treatment group hazard rate.  The last three bins were combined for the +0m treatment group, and the last two bins were combined for the +24m treatment group. 
%The effective number of parameters used in the calculation of BIC and AIC for the piece-wise exponential models is equal to the sum of the $J$ for each of the two hazard rates, plus 7 (the number of fixed covariates in the model).

\subsubsection{Comparison of Estimated Hazard Ratio and Predictor Effects}\label{sec:parcomp}
\begin{figure}[htbp] %%%% Comparing betas, PH
	\centering
	\includegraphics[width=5in]{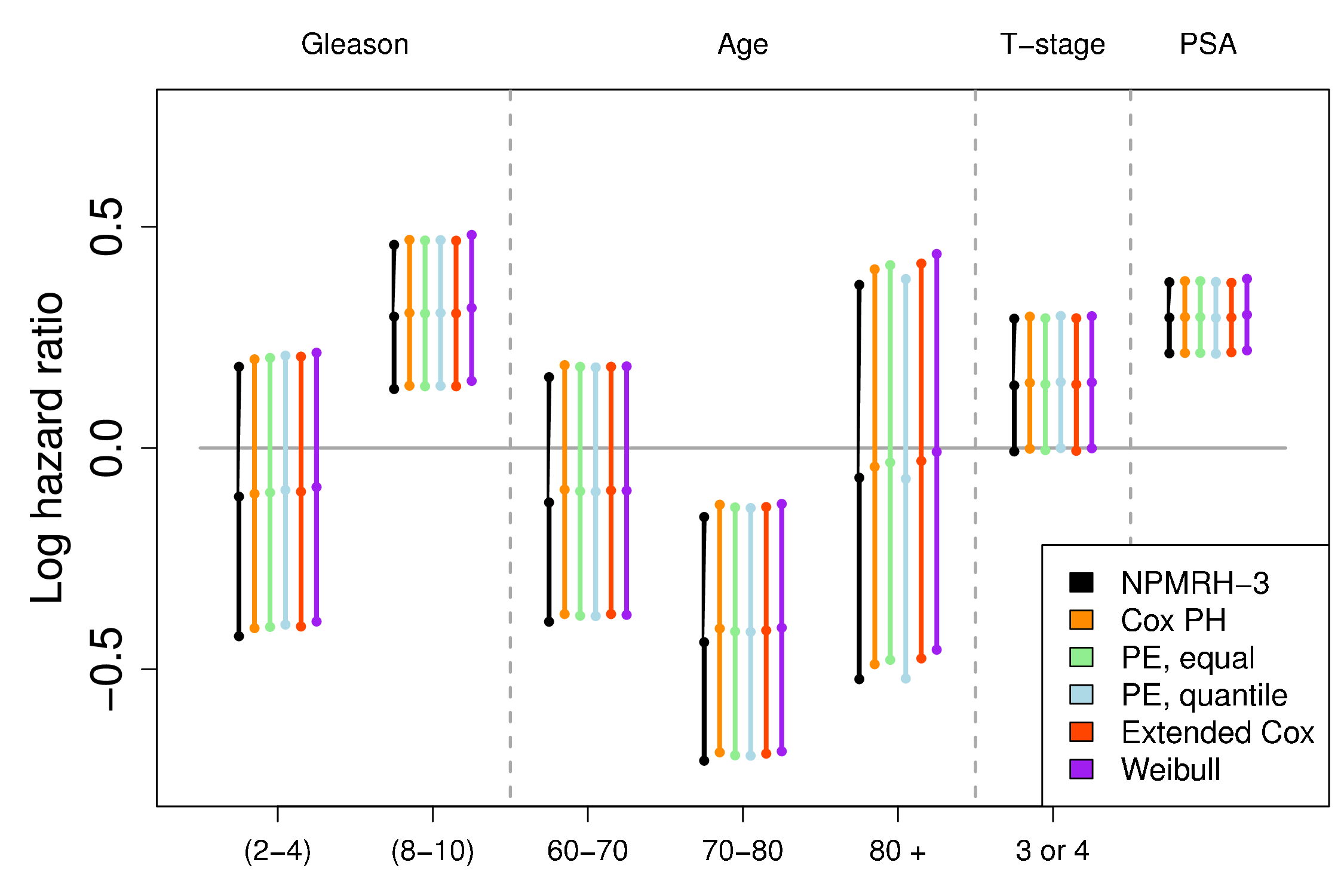}
	\caption{\footnotesize The estimates for the proportional hazards covariate effects and corresponding 95\% intervals (credible intervals for NPMRH-3, confidence intervals for the remaining models). Differences are minor, even among the widths of the 95\% intervals, between the different models.  The AFT and DDP survival model estimates are not included as their number and interpretation is different than the standard Cox PH covariate interpretation.}
	\label{fig:logratPH}
\end{figure}

\begin{figure}[htbp] %%%% Comparing betas, NPH
	\centering
	\includegraphics[width=6in]{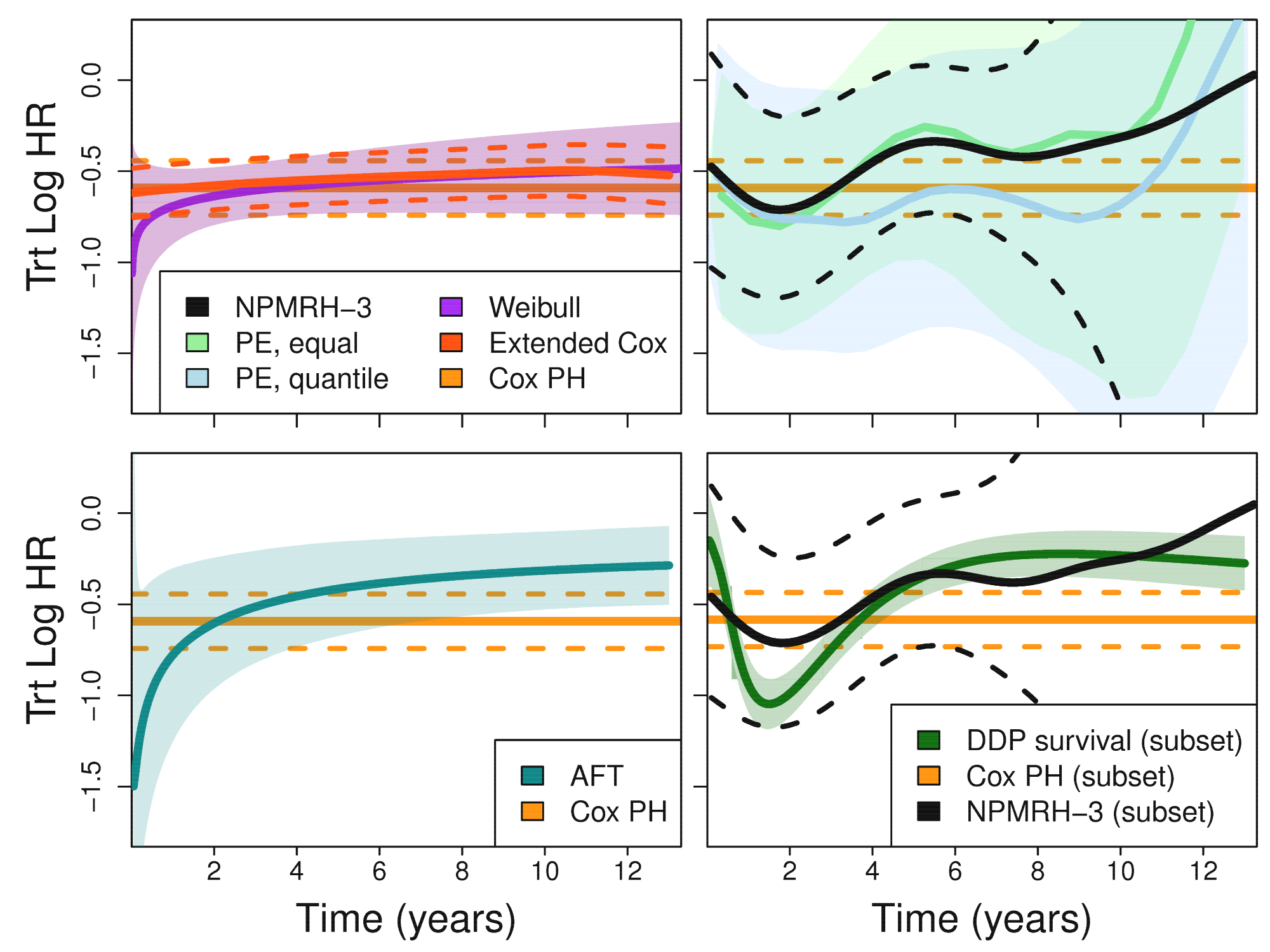}
	\caption{\footnotesize Comparison of the estimated treatment log-HR from the different models.  TOP: The estimated log-hazard ratio of the +24m AD therapy effects over time, with 95\%  bounds (smoothed point-wise credible intervals for the MRH model, and point-wise confidence intervals for the remaining models).  In the top left graph, it can be observed that after the initial dip in the NPH Weibull model estimate, the log-ratio slowly but steadily increases over the course of the study.  The 95\% confidence interval bounds for the log-HR of the NPH Weibull model are much narrower than most of the other models, which is expected due to dramatically fewer parameters estimated in that model.  The extended Cox model shows the same upward trajectory, although the initial dip is not pronounced, which is likely due to the choice of the smoothing parameter.  The top right graph shows similarities between the log-HR estimates for the MRH and adjusted PE models, although the PE models have a sharper upward trend towards the very end of the study.  BOTTOM: The estimated log-hazard ratio of the +24m AD therapy effects over time, with 95\% bounds (smoothed point-wise credible intervals for the MRH model, and approximated point-wise confidence intervals for the AFT and DDP survival models).  On the left, the AFT model shows a similar pattern to the NPH Weibull model, with an initial dip at the beginning of the study, followed by an increasing estimate over time.  On the right, the DDP survival model (calculated on the subset of significant predictors), is contrasted against the NPMRH-3 model (also calculated on the subset of significant predictors), where they show a similar pattern, with an initial log-HR estimate greater than the Cox PH estimate, followed by a dip at 2 years.  Unlike the other models, the DDP survival model estimated log-HR treatment effect decreases slightly towards the end of the study.  For contrast, the Cox PH model estimated treatment effect and 95\% confidence interval is shown in all figures.  All models have periods where the estimated treatment log-HR extends outside the 95\% confidence interval for the Cox PH model treatment effect.  The estimated hazard rate and 95\% credible interval bounds for the AFT model were predicted for all covariate groups using the \texttt{bayessurvreg1()} function found in the ``bayesSurv" package in R \citep{bayesSurvPkg}, and were predicted for the DDP survival model using the \texttt{LDDPsurvival()} function in the ``DPpackage" package in R \citep{DPpkg}.  In both models, the bounds were used to approximate the point-wise variance of the hazard rates, which were then used to approximate the point-wise variance of the log-HR. }
	\label{fig:logratNPH}
\end{figure}

The estimates from all models are compared visually in Figures \ref{fig:logratNPH} and \ref{fig:logratPH}.  All models were remarkably similar in terms of the proportional hazard covariate effects, both in  point estimates and their  95\%  bounds (credible intervals for the MRH model, and confidence intervals for the remaining models), as can be seen in Figure \ref{fig:logratPH}.  Given this similarity,  we refer the readers to  Table~\ref{tab:mrhres} and Section~\ref{section:MRHres} for  interpretation and discussion of these effects in the NPMRH-3 model.  (Note that the estimated covariate effects for the AFT and DDP survival models are not shown as their number and interpretation are different than the other models.)  

In contrast, the estimates of the time-varying treatment effect show notable differences  (Figure \ref{fig:logratNPH}).  The PE and MRH models provide similar estimates, although the PE model log-HR estimates exhibit a rapid increase towards the end of the study when the number of observed failures becomes sparse (top right graph).  Due to its parametric form, the NPH Weibull model has an initial dip in the estimated treatment effect, and then slowly but steadily increases over the course of the study, although the estimated effects remain negative throughout the study period.  The extended Cox model shows a similar pattern and estimate, without the initial dip (top left graph).  The AFT estimated log-HR follows a trajectory similar to that of the NPH Weibull model, including the wider 95\% confidence interval bounds in the initial study period (bottom left graph).  The DDP survival model (calculated on the subset of significant predictors only) shows an initial pattern similar to that of the NPMRH-3 model (also calculated on the subset of significant predictors only), with a dip at the two year mark, followed by an upward trend.  The DDP survival model is the only model that shows a  possible decreasing trend towards the end of the study.  For comparison, the Cox PH model treatment estimate (included under the proportional hazards assumption), has been included in all graphs as a constant value over time ($\hat{\beta} = -0.59$, 95\% CI: -0.74, -0.44).   It can be observed that throughout various periods of the study, all models have estimated treatment effects that extend outside the 95\% confidence interval for the Cox PH model treatment effect.  In addition, all models show that long-term treatment is beneficial over longer periods of time, even if the effects may be diminishing.  

\subsubsection{Sensitivity Analysis to Parameter $k$ in the MRH Models}\label{sec:investigatek}

In all the MRH models, the parameter $k$ controls the correlation among the hazard increments within each bin (\cite{Bouman}, see Appendix A for details).  The default value for $k$ in the above analyses was 0.5, which implies zero {\it a priori} correlation among the hazard increments.  However, when $k > 0.5$, the increments are positively correlated \textit{a priori}, and, similarly, when $k < 0.5,$ the hazard increments are negatively correlated \textit{a priori}.  Another way to understand the impact of $k$ is that higher values lead to smoother hazard functions. 

In practice, different approaches to choosing a hyperprior for $k$, including empirical Bayes methods, are possible. However,  $k$ will in general tend to depend on  the resolution level \citep{Bouman}, as well as with the significance level used in the pruning algorithm \citep{Yprune}. Both the resolution and the pruning can be also used to imply the desired {\it a priori} level of smoothness of the hazard function. For this reason, we fix $k$ in the above analyses, and  perform a sensitivity analysis to  examine the effect the choice of k might have on the posterior hazard rate  estimates. We only examine the effects of different values of $k$ in the 3-level pruned MRH model (NPMRH-3) (see  Subsection~\ref{sec:gofsec} for motivation.) 

\begin{figure}[htbp] %%%% MRH log-ratio estimates, compare k
	\centering
	\includegraphics[width=5.5in]{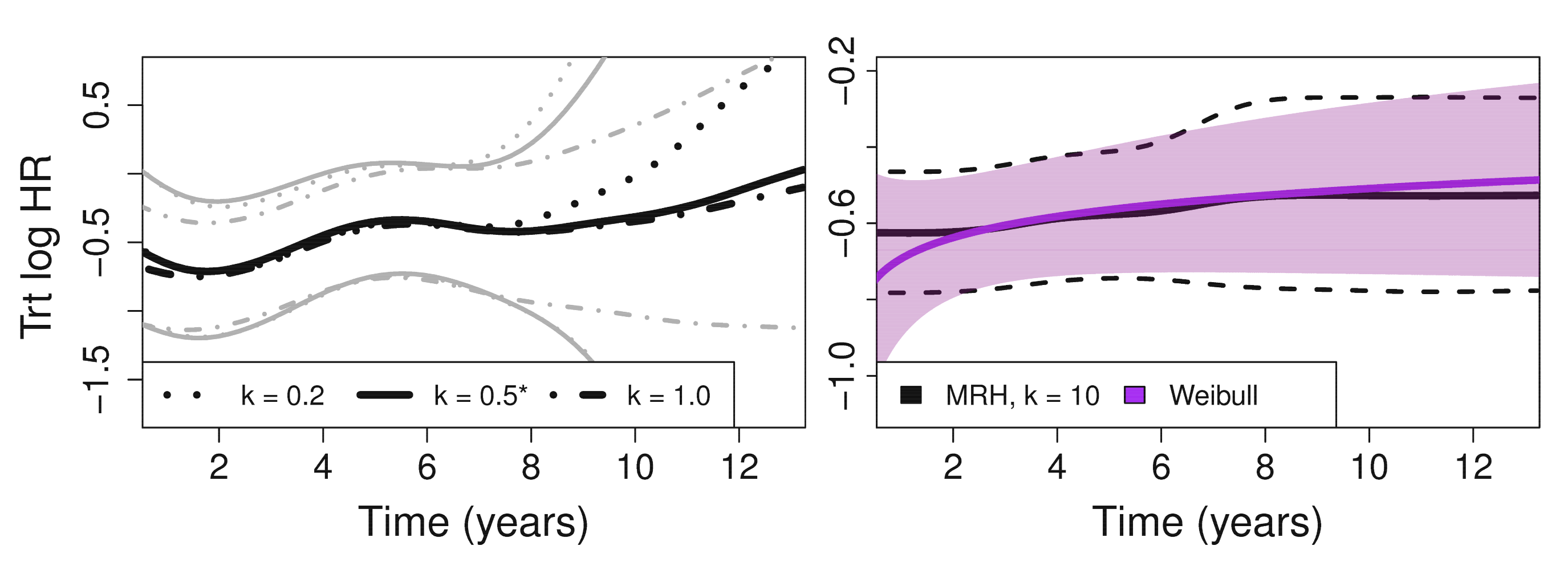}
	\caption{\footnotesize LEFT: Comparison of the smoothed, estimated log-hazard ratios for the NPMRH-3 model with $k$ values equal to 0.2, 0.5 (the default MRH model, denoted with a `*'), and 1.0, with estimates shown in black and smoothed point-wise 95\% credible interval bounds shown in grey.  The model with $k = 0.2$ is the least smooth, with the largest 95\% credible interval bounds.  Conversely, the model with $k = 1$ is  smoother, with narrower 95\% credible interval bounds, as the positive correlation allows more ``shared" information between the hazard increments.  RIGHT: The estimated log-hazard ratio for the NPMRH-3 model with $k = 10$ contrasted with the  NPH Weibull model from the previous section.  This figure highlights the adaptability of the MRH model; if $k$ is fixed at a large value (making the hazard rate quite smooth), the NPMRH-3 model closely mimics the results of the Weibull NPH model.  Note that the two figures do not have the same y-axis scale.}
	\label{fig:logratios_ks}
\end{figure}

The sensitivity analysis results are displayed in Figure  \ref{fig:logratios_ks}. On the left plot in Figure  \ref{fig:logratios_ks}, the original NPMRH-3 model (with $k = 0.5$) is contrasted against the models with negatively correlated hazard increments ($k = 0.2$), and positively correlated hazard increments ($k = 1.0$).  As anticipated,  in the negatively correlated model the log-HR is less smooth,  and has wider 95\% credible intervals, resembling the PE model results.  However, the NPMRH-3 model with $k = 1.0$ is smoother, with narrower 95\% credible intervals.  The positive correlation between hazard increments results in smoother posterior estimates,  as more information is shared across bins.  The right graph of Figure \ref{fig:logratios_ks} highlights the adaptability of the MRH model in controlling the smoothness of the log-hazard ratio through the parameter $k$.  In this instance, with $k$ fixed at a very high value of 10 (highly positively correlated increments), the NPMRH-3 model closely mimics the results of the parametric NPH Weibull  model. 

\subsubsection{Model Performance Comparison}\label{sec:gofsec}
In addition to model parameter comparisons in Subsection~\ref{sec:parcomp}, the set of  models were also compared based on their  goodness of fit (GOF), as well as several information criteria. The GOF was evaluated using the following simple measure over time:
\begin{eqnarray*}
	GOF(t) =\frac{1}{n_t} \sum_{i = 1}^{n_t} |I_i\{\text{biochemical failure occurs $> t$}\} - P(\text{subject $i$ experiences biochemical failure $> t$})|,
\end{eqnarray*}
where $|\cdot|$ denotes  absolute value, $I_i\{\text{biochemical failure occurs $> t$}\}$ is an indicator variable which equals $1$ if the subject $i$ fails after time $t$ and equals 0 otherwise, and $P(\text{subject $i$ experiences biochemical failure $> t$})$ is the model-based probability of the subject $i$ experiencing biochemical failure.   This probability is found based on the estimated model parameters (posterior medians, or maximum likelihood estimates) and covariates for subject $i$. Patients who were censored before time $t$ were not included in the GOF calculation at time $t$.  Therefore, $n_t$ represents the total number of patients in the cohort minus the number of patients censored before time $t$, so  that the maximum value the GOF statistic can take is 1. In other words, the GOF measure calculates the average difference between the observed failure time and the probability of failure at that time point. Lower GOF values indicate more accurate failure approximations and a better fitting model.  

\begin{figure} %%%% GOF graph
	\centering
	\includegraphics[width=6.5in]{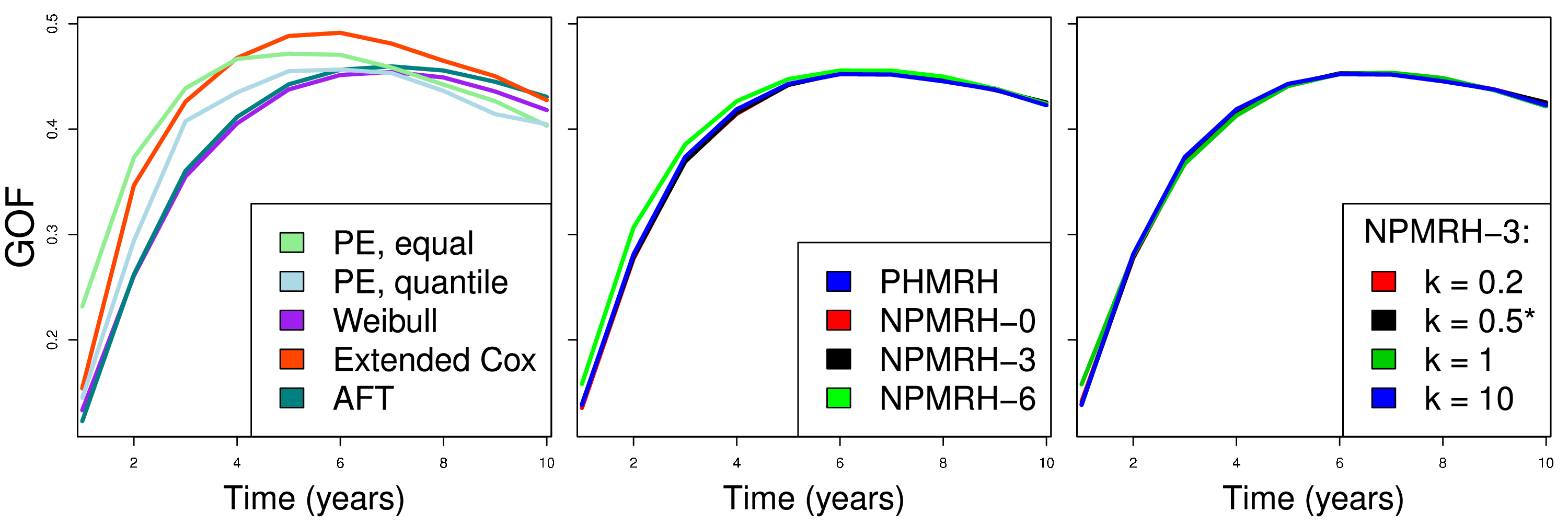}
	\caption{\footnotesize LEFT: Goodness-of-fit values across time for both adjusted PE models, the NPH Weibull model, the extended Cox model, and the AFT model.  CENTER: Goodness-of-fit values across time for the four MRH models (PHMRH, NPMRH-0, NPMRH-3, and NPMRH-6).  RIGHT:  Goodness-of-fit values across time for the NPMRH-3 model with different fixed $k$ values equal to 0.2, 0.5, 1.0 and 10.  The `*' placed by $k = 0.5$ denotes that this is the original NPMRH-3 model reported above.  There are few differences among most models,  with almost indistinguishable differences between the MRH model with all levels pruned and the MRH model with only 3 levels pruned or the MRH models with different values of $k$. (Details on the different values of $k$ in the NPMRH models are discussed in section \ref{sec:investigatek}.) However, the extended Cox model and both PE models have higher GOF values for the first seven years when compared to the others. Note that the GOF statistic was not calculated for the Cox PH model as no estimated hazard rate was available, and was also not calculated for the DDP survival model, as we did not have access to the subject-specific survival curves.}
	\label{fig:GOF}
\end{figure}

Results from the GOF statistic calculations are shown in Figure \ref{fig:GOF}.  Most models show very similar results and trajectories, with exception for both of the adjusted PE models and the extended Cox model.  The adjusted PE model with equal bins has the worst survival prediction initially, followed by the extended Cox model and the adjusted PE model with bins determined through quantiles.  After four years, the extended Cox model has the highest GOF of all models.  Differences between the MRH models (including those with different values of the prior parameter $k$) are negligible, and also very similar to the results for the NPH Weibull and AFT models.  Note that the GOF statistic was not calculated for the Cox proportional hazards model, as no estimate of the hazard rate is typically produced by Cox models.  That statistic was also not calculated for the DDP survival model, as subject-specific survival curves are not provided in that package.

Table~\ref{tab:DICtable} shows several information criteria (AIC, BIC, and DIC where appropriate) for all the models considered (with the exception of the DDP survival model, as subject-specific hazard rates and survival curves are not provided in that package). Among the MRH models, the PHMRH model has the highest DIC value, which is about 5000 points greater than any of the NPMRH models.  It also has the highest negative log-likelihood, BIC and AIC values, despite the smaller number of parameters when compared to the NPMRH models.  This is consistent with our earlier observation that the PH model does not seem to provide a good description of  the data.  

When comparing the NPMRH models with different levels of pruning (NPMRH-0, NPMRH-3, NPMRH-6), the NPMRH-3 model has the lowest negative log-likelihood value, followed closely by the NPMRH-6 model. The NPMRH-6 model  has the  lowest DIC, BIC, and AIC values as it has the smallest number of estimated parameters of all MRH models considered.  However, all three NPMRH models have very similar information criteria values, with the exception of BIC for NPMRH-0 whose penalty for its large number of parameters sets it apart from the rest of the models.  It is also notable that among the NPMRH-3 models, the lowest negative log-likelihood, DIC, BIC, and AIC values are for the model with $k = 0.2$, which may be a good choice for examining the hazard rate of biochemical failure for this particular data set, as it captures the most details in the failure pattern.  The negative log-likelihood values (and hence BIC and AIC calculations) of the adjusted PE models are slightly smaller than those of the NPMRH models, although the values are comparable.  When compared to the NPH Weibull models, the NPMRH models all have lower negative log-likelihood values.  However, BIC and AIC values are higher in the NPMRH models due to the higher number of estimated parameters.  The AFT model has a higher negative log-likelihood value when compared to the other models (with the exception of the PHMRH model), and the extended Cox model has a slightly higher negative log-likelihood value when compared to the MRH models, but the values are similar. Regardless of model choice however, all evidence points to the treatment effects  not being proportional:  the effects of an additional 24 months of AD therapy   change over the entire length of the study.

\begin{table} %% DIC table to compare models
	\center
	\begin{tabular}{|cl||c|c|c|c|c|}
		\hline
		\multicolumn{2}{|c||}{\multirow{2}{*}{Model}}&\multirow{2}{*}{-2*log(L)}&Effective Number&\multirow{2}{*}{DIC}&\multirow{2}{*}{BIC}&\multirow{2}{*}{AIC}\\
		&&&of Parameters&&&\\
		\hline\hline
		\multirow{8}{*}{MRH}&\multicolumn{1}{|l||}{PHMRH}&12628.0&32&9651.1&12860.3&10555.8\\
		\cline{2-7}
		&\multicolumn{1}{|l||}{NPMRH-0}&4703.1&139&4751.5&5712.1&4981.1\\
		&\multicolumn{1}{|l||}{NPMRH-3}&4669.8&43&4665.0&4981.9&4755.8\\
		&\multicolumn{1}{|l||}{NPMRH-6}&4679.0&38&4602.1&4954.8&4755.0\\
		\cline{2-7}
		&\multicolumn{1}{|l||}{NPMRH-3 ($k = 0.2$)}&4667.9&43&3582.9&4980.1&4753.9\\
		&\multicolumn{1}{|l||}{NPMRH-3 ($k = 0.5*$)}&4669.8&43&4665.0&4981.9&4755.8\\
		&\multicolumn{1}{|l||}{NPMRH-3 ($k = 1.0$)}&4700.7&43&4298.7&5012.9&4786.7\\
		&\multicolumn{1}{|l||}{NPMRH-3 ($k = 10$)}&4792.8&43&4578.2&5105.0&4878.8\\
		\hline
		\multirow{2}{*}{PE}&\multicolumn{1}{|l||}{Equal bins}&4611.1&42&-&4916.0&4695.1\\
		&\multicolumn{1}{|l||}{Quantile bins}&4596.7&56&-&5003.2&4708.7\\
		\hline
		\multicolumn{2}{|l||}{NPH Weibull}&4759.9&11&-&4839.7&4781.9\\
		\hline
		\multicolumn{2}{|l||}{AFT}&5277.9&-&-&-&-\\
		\hline
		\multicolumn{2}{|l||}{Extended Cox}&4747.0&-&-&-&-\\
		\hline
	\end{tabular}
	\caption{\footnotesize Information criteria (DIC-when applicable, BIC, and AIC) for the 4 MRH models (PHMRH, NPMRH-0, NPMRH-3, and NPMRH-6), the NPHMRH-3 model with different fixed values of $k$ (in the $R_{m,p}$ prior),  the non-proportional hazards (NPH) Weibull model, the adjusted piece-wise exponential models (adjusted by allowing bins to be merged), the AFT model, and the smoothed extended Cox model.  In addition, the values of twice the negative log-likelihood ($-2*log(L)$) and the effective number of parameters (when known) are shown.  Lower DIC, BIC, and AIC values represent  models better supported by the data. Details on the different values for $k$ in the NPMRH models will be discussed in section \ref{sec:investigatek}.  The DDP survival model is not included as the log-likelihood and information criteria values are not available from the fitted model.}
	\label{tab:DICtable} 
\end{table}

%%%%%%%%%%%%%%%%%%%%%%%%%%%%%%%%%%%%%%%%%%%% %%%%%%%%%%%%%%%%%%%%%%  
%											 Discussion 
%%%%%%%%%%%%%%%%%%%%%%%%%%%%%%%%%%%%%%%%%%%% %%%%%%%%%%%%%%%%%%%%%%   
\section{Discussion}

This paper illustrates how different modeling and smoothing assumptions effect the estimate of the time-varying treatment effect.  We present results from a suite of models ranging from parametric to non-parametric, and demonstrate that different assumptions can lead to very smooth, flat log-hazard ratio estimates (such as those in the NPH Weibull model) to estimates which vary more over time (such as those in the MRH, PE, and the DDP survival model).  Additionally, the different models exhibited a high degree of variability in the goodness-of-fit measure and the penalized goodness of fit criteria.  We have also shown how  choosing different values of $k$ gives the MRH model the flexibility to perform similarly to other models, ranging from the piece-wise exponential to the parametric Weibull model.  The NPMRH model allows for multiple changes in the treatment effects over time, with multiple increases and decreases  over the length of a study period. 

Other patient and disease characteristic covariate effects were similar to those previously seen in this trial \citep{Horwitz} and expected based on the effects of these factors in other studies. Men with higher Gleason scores had greater hazard of biochemical failure, although this difference was statistically significant only for those with Gleason scores of 8 or more. In addition, those with more advanced tumor stage (T-stage 3 or 4) or with higher PSA level at diagnosis also had a higher hazard rate of biochemical failure.  Men who were older at diagnosis were found to have a lower hazard rate of biochemical failure, although this may be still partly  confounded with the censoring patterns in older patients and warrants further exploration.

Additionally, the presented analysis has allowed insight into the effects of the duration of AD therapy on biochemical failure, and in particular into  how the effects of AD therapy changed  throughout the course of the study.  While it was already apparent that 24 months of additional AD therapy is beneficial (relative to the 0 additional months of AD therapy) in that it prolongs the time until biochemical failure and other failure endpoints \citep{Horwitz}, our investigation has revealed additional insights. During and immediately after active therapy, the peak in the hazard rate around two years is much flatter for the +24m treatment group.  In addition, the +24 month group continued to have a lower hazard rate throughout most of the observation period (over 10 years), although smaller due to the non-proportionality of the treatment effect. Thus, it does appear that the benefits of the additional months of AD therapy,  while diminishing over time, are persistent, which suggests that failure in the longer AD duration group are not simply deferred but possibly avoided. On the other hand, for those patients who received short AD therapy and did not fail early or during the peak period of failures, their late term prognosis is nearly as favorable as those who underwent long duration AD.  Thus, until such  patients can be prospectively identified, the long AD approach would seem to be preferred for all patients. To this end, we also illustrate how the Bayesian approach  can allow the use of posterior predictive failure probabilities, such as in Figure \ref{fig:survCurveComparison}, as aids in clinical contexts. 

\section*{Appendix A: Details on the MRH prior and Pruning Method}
\subsection*{MRH prior}
The foundation of the MRH method is a tree-like wavelet-based multi-resolution prior on the hazard function, chosen conveniently to allow  scalability and consistency across different time scales (i.e minutes, weeks, years, etc). It uses a piece-wise constant approximation of the hazard function over    $J$  time intervals, parametrized by a set of  hazard increments  $d_{j}, j = 1,\dots, J$. Here, each $d_j$ represents the aggregated hazard rate over the $j^{th}$ time interval, ranging from $(t_{j-1}, t_j)$.  In the standard survival analysis notation,  $d_{j} = \int_{t_{j-1}}^{t_{j}} h(s)ds \equiv H(t_{j})-H(t_{j-1})$, where $h(t)$ is the hazard rate at time $t$.  

\begin{figure}% MRH graphic
	\centering
	\includegraphics[width=5in]{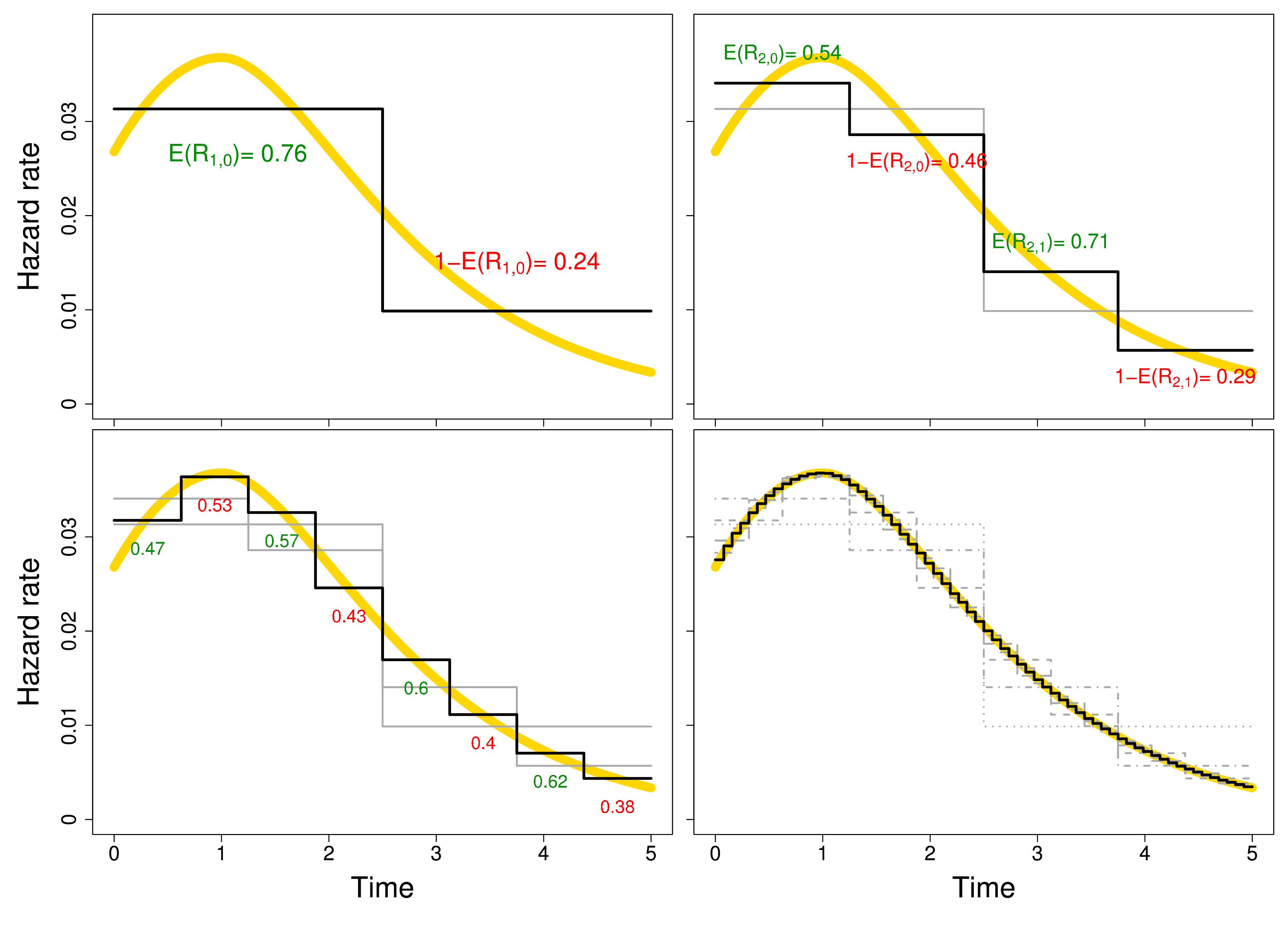}
	\caption{\footnotesize  Example of an MRH prior mean (black), centered at a desired parametric hazard rate (yellow), at various  resolution levels $m$.  The first figure (upper left) shows the mean MRH  rate at the first level ($m = 1$), with $E(H_{1,0}) = E(H)*E(R_{1,0})$ and $E(H_{1,1}) = E(H)*E(1-R_{1,0})$. 
$R_{1,0}$ prior mean of 0.76  reflects the higher hazard during the first time interval.  The second figure (upper right) shows the mean MRH rate at the second level ($m = 2$), with $E(H_{2,0}) = E(H)*E(R_{1,0})*E(R_{2,0})$, and $E(H_{2,1}), E(H_{2,2})$ and $E(H_{2,3})$ derived analogously.  The third figure (lower left), shows the mean MRH rate at the third level ($m = 3$).  The last figure (lower right), shows the mean MRH  rate for $m = 6$, which closely matches the true hazard rate.  In all the figures, the mean MRH rates for the previous resolutions are shown in grey,  $E(R_{m,p})$ are shown in green, and $E(1-R_{m,p})$  in red. The advantage of the MRH model is its ``self-consistency" under aggregation \citep{Yprune}, which means that the prior specification at any level $m$ is independent of the ultimate level $M$. This property allows the hazard rate to be examined at multiple resolutions in a consistent manner.}
\label{fig:MRHfig}
\end{figure}

To facilitate the recursive diadic partition of the multiresolution tree, we assume that $J=2^{M}$. Here, $M$ is an integer, set large enough to achieve the desired  time resolution for the hazard rate. $M$ can also be chosen using model selection criteria or clinical input, as in  \cite{Bouman}, and \cite{Dignam}.    Note that the cumulative hazard, $H$, is equal to the sum of all  $2^M$ hazard increments $d_{j}, j = 1,\dots2^M$. The model then recursively splits $H$ at different branches via the ``split parameters" $R_{m,p} = H_{m,2p}/H_{m-1,p}, \,\, m = 1,2,\dots,M-1, \,\, p = 0,\dots,2^{m-1}-1$. Here, $H_{m,q}$ is recursively defined as $H_{m,q}\equiv H_{m+1,2q}+H_{m+1,2q+1}$ (with $H_{0,0} \equiv H$, and $q = 0,\dots,2^{m}-1$).  The $R_{m,p}$ split parameters, each between 0 and 1,  guide the shape of the {\it a priori} hazard rate over time (Figure \ref{fig:MRHfig}).

The complete hazard rate prior specification  is obtained via priors placed on all tree parameters:  a Gamma($a, \lambda$) prior is placed on the cumulative hazard $H$, and Beta prior on each split parameter $R_{m,p}$,  $\mathcal{B}e(2\gamma_{m,p}k^{m}a,2(1-\gamma_{m,p})k^{m}a)$.  For example, the priors for $H$ and $R_{m,p}$ in 3-level MRH model ($M=3, J=8$) would be:
\begin{eqnarray}\label{mrprior_start}
H  \sim & \ \mathcal{G}a(a,\lambda),   \nonumber \\ \nonumber
R_{1,0}  \sim & \ \mathcal{B}e(2\gamma_{1,0}ka,2(1-\gamma_{1,0})ka),\\ \nonumber
R_{2,p} \sim  & \ \mathcal{B}e(2\gamma_{2,p}k^{2}a,2(1-\gamma_{2,p})k^{2}a),~ p = 0,1\\
R_{3,p} \sim  & \ \mathcal{B}e(2\gamma_{3,p}k^{3}a,2(1-\gamma_{3,p})k^{3}a),~ p= 0,1,2,3.\label{mrprior_end}
\end{eqnarray}
Under this parametrization,   the prior distribution of each hazard increment is governed by  these Beta and Gamma distributions. In particular, their prior expectations depend on the hyperparameters of the Beta and Gamma  priors -- for example, in the above 3-level model $E(d_1) = E(H) E(R_{1,0}) E(R_{2,0}) E(R_{3,0})$. Similarly, these MRH  hyperparameters control the correlation between the hazard increments $d_j$, and thus directly relate to the smoothness of the multiresolution prior, as shown in \cite{Bouman} and \cite{Yprune}. This parametrization also insures the self-consistency of the MRH prior at multiple resolutions \citep{Bouman, Yprune}.

\subsection*{Pruning the MRH tree} 

The  MRH prior resolution is chosen as a compromise between the desire for detail in the hazard rate, and the amount of data. As  the resolution increases (and the number of time intervals increases), observed failure counts within each bin will decrease. While useful for revealing detailed patterns, a large number of  intervals (and consequently,  a large number of model parameters) will generally require longer computing times and  result in estimators with lower statistical efficiency \citep{Yprune}. ``Pruning'', as used in \citet{Yprune}, is a data-driven pre-processing technique, which  combines consecutive $H_{m,p}$s that are statistically similar (and happens frequently with periods of low failure counts).  The technique increases the computational efficiency by decreasing the parameter dimension \textit{a priori}, which can greatly speed up  analyses of non-proportional hazards.  The pruning method  thus changes the overall time resolution of the MRH prior, keeping the higher resolution during the periods of high event counts, and lower resolution during periods of low event counts. 

The  MRH pruning technique has been extensively studied in \cite{Yprune}.  Briefly, pruning starts with the full MRH tree prior, and merges adjacent bins that are constructed via the same split parameter, $R_{m,p}$,  when the hazard increments in these two bins ($H_{m+1,2p}$ and $H_{m+1,2p+1}$) are statistically similar.
This is  inferred by testing the hypothesis $H_0: R_{m,p}=0.5$  against the alternative $H_a: R_{m,p} \neq 0.5$, with a pre-set type I error $\alpha$,  for each split parameter $R_{m,p}$ ($p = 0,...,2^{m-1}-1$). If the null hypothesis is not rejected,  that split $R_{m,p}$ is set to $0.5$ and the adjacent hazard increments are considered equal and the time bins declared ``fused''.  The hypothesis testing can be applied  to all $M$ levels of the tree or just a higher resolution subset of the tree.  While the pruning is expected to reduce the amount hazard rate detail discovered by the MRH method, the posterior hazard rate estimator is shown to have lower risk compared to its equivalent from the non-pruned model \citep{Yprune}.

\section*{Appendix B: Estimation and Pruning Steps}
The estimation algorithm is  performed two steps: the pruning step and the Gibbs sampler routine.  The details are listed below.

\subsection*{Pruning step}
The pruning step is run only once for each of the $\mathcal{L}$ hazard rates at the beginning of the algorithm as a pre-processing step in order to finalize the MRH tree priors. The $R_{m,p; \ell}$ parameters for which the null hypothesis is not rejected are set to $0.5$ with probability 1, while the rest are estimated in the Markov chain Monte Carlo (MCMC) routine.

%%%%%%%%%%%%%%%%% GIBBS
\subsection*{Gibbs sampler steps}

After the pruning step, the Gibbs sampler algorithm is performed to obtain the approximate posterior distribution of $H_\ell, a_\ell, \lambda_\ell, k_\ell, \gamma_\ell$, and the $R_{m,p; \ell}$ that have not been set to 0.5 for each stratum ($\ell = 1,\dots, \mathcal{L}$) as well as $\vec \beta$.  

The algorithm is as follows, with steps repeated until convergence:

\begin{enumerate}
\item For each of the $\mathcal{L}$ treatment hazard rates ($\ell = 1, \dots, \mathcal{L}$):
\begin{enumerate}
\item \label{item:Gstep1}%% SAMPLE H 
Sample $H_\ell$ from the posterior for $H_\ell$, which is  a gamma density with the shape parameter
$a_\ell + \sum_{i \in \text{tx}_\ell}\delta_{i, \ell}$, and rate parameter $\lambda_\ell^{-1} + \sum_{i \in \text{tx}_\ell} \mrm{exp}\left(X'_{i, \ell}\vec\beta\right)F_\ell(T_{i, \ell})$, where  $F_\ell(T_{i, \ell})$ = $H_\ell(\min(T_{i, \ell},t_{J}))/H_\ell(t_{J})$.
	
\item %% SAMPLE A, LAMBDA
Sample $a_\ell, \lambda_\ell$ from their respective posterior distributions (see below).

\item %% SAMPLE RMP
Sample each $R_{m,p; \ell}$ for which the null hypothesis was rejected from the full conditional:
\begin{equation*}\label{Rpost}
\begin{array}{l}
{ R_{m,p; \ell }^{2\gamma_{m,p; \ell} k_\ell^m  a_\ell -1}(1-R_{m,p; \ell })^{2(1-\gamma_{m,p; \ell} )k_\ell^m  a_\ell -1} }
\Pi_{i\in \text{tx}_\ell}\left\{\left[h_{\ell 0}\left(T_{i, \ell}\right)\right]^{\delta_{i, \ell}}\mrm{exp}\left(-\mrm{exp}\left(X'_{i, \ell}\vec\beta\right)H_\ell F_\ell(T_{i, \ell})\right)\right\}.
\end{array}
\end{equation*}

\item \label{item:GsteplastL}%% SAMPLE K, GAMMA
Sample $k_\ell, \gamma_{m,p; \ell}$ from their respective posterior distributions (see below).
\end{enumerate}

\item \label{item:Gsteplast}%% SAMPLE BETA
With a $\mathcal{N}(0,\sigma_{\beta_s}^2)$ prior (with a known  variance) on  each covariate effect  modeled under the proportional hazards assumption,  $\beta_s \,\, (s = 1, \dots, z$), each has the following full conditional distribution:
 \begin{equation*}
 \begin{array}{l}
\pi( \beta_s | \beta_s^{-}) \propto  \ds 
\left(\Pi_{\ell = 1}^\mathcal{L} \Pi_{i\in \text{tx}_\ell} \big[\mrm{exp}\left\{X_{ij, \ell}\beta_s\right\}\big]^{\delta_{i, \ell}} \ds \mrm{exp}\left\{-\mrm{exp}\left(X'_{i, \ell}\vec\beta\right)H_\ell F_\ell(T_{i, \ell})\right\}\right)\mrm{exp}\left\{-\frac{\beta_s^2}{2\sigma_{\beta_s}^2}\right\}
\end{array}
\end{equation*}
Note that this posterior distribution includes the full set of observations and covariates, from all  strata jointly.
\end{enumerate}

\subsection*{Full conditionals for the hyperparameters  $a, \lambda, k$, and $\gamma_{m, p}$}
The parameters in the prior distributions of $H$ and all $R_{m,p}$s for each covariate stratum  ($\ell = 1,\dots,\mathcal{L}$),  $a_\ell, \lambda_\ell, k_\ell$, and $\gamma_{\ell ; m, p}$, can either be fixed at desired values, or treated as random variables with their own set of hyperpriors.   In the case of the latter, they would be sampled within the Gibbs sampler separately for each stratum, according to their own full conditional distributions.  Below are the forms of these full conditional distributions for a specific set of hyperpriors we chose.

For notational simplicity, the stratum-specific index is suppressed  below. The notation $\eta^-$ will be used to denote the set of all data and all parameters except for the parameter $\eta$ itself.  The full conditionals are as follows:
\begin{itemize}

\item If $a$ is given a zero-truncated Poisson prior, $\ds \frac{e^{-\mu_{a }}\mu_{a }^{a }}{a !\left(1-e^{-\mu_{a }}\right)}$ (chosen for computational convenience), the full conditional distribution for $a $ is:
\begin{equation*}
\begin{array}{l}
\ds \pi (a   \mid a ^{-}) \propto \ds \frac{H ^{a }\mu_{a }^{a }}{\lambda ^{a }(a -1)! a  !} \frac{\ds }{}
  \Pi_{m=1}^{M}\Pi_{p=0}^{2^{m-1}-1}\left\{\frac{ R_{ m,p }^{2\gamma_{ m,p} k ^m  a  }(1-R_{ m,p })^{2(1-\gamma_{ m,p} )k ^m  a  } }{\mrm{B}(2\gamma_{ m,p} k ^m  a  ,2(1-\gamma_{ m,p} )k ^m  a  )}\right\}
 \end{array}
 \label{eq:apost}
\end{equation*}

\item If the scale parameter $\lambda$ in the gamma prior for the cumulative hazard function $H$ is given an exponential prior with mean $\mu_{\lambda}$, the resulting full conditional is:
\begin{equation*}
\ds \pi(\lambda  |\lambda^{-})\propto \ds \frac{1}{\lambda^{a}}\exp\left\{-\left(\frac{ H}{\lambda}+\frac{\lambda}{\mu_{\lambda}}\right)\right\}
\end{equation*}

\item If  $k$ is given an exponential prior distribution with mean $\mu_{k}$,  the full conditional distribution for $k$ is  as follows:
\begin{equation*}
\begin{array}{l}
\ds \pi(k  \mid k^{-})\propto \Pi_{m=1}^{M}\Pi_{p=0}^{2^{m-1}-1}
 \ds \left\{\frac{ R_{m,p }^{2\gamma_{m,p}k^m  a }(1-R_{m,p })^{2(1-\gamma_{m,p})k^m  a } }{\mrm{B}(2\gamma_{m,p}k^m  a ,2(1-\gamma_{m,p})k^m  a )}\right\}
e^{-\frac{k}{\mu_{k}}}
\end{array}
\label{eq:kpost}
\end{equation*}

\item If a Beta($u$, $w$) prior is placed on each $\gamma_{m,p}$, the full conditional distribution for each $\gamma_{m,p}$ is proportional to:
\begin{equation*}
\ds
   \frac{ R_{m,p }^{2\gamma_{m,p} k^m  a }(1-R_{m,p })^{2(1-\gamma_{m,p})k^m  a } }{\mrm{B}(2\gamma_{m,p} k^m  a ,2(1-\gamma_{m,p})k^m  a )}
 \gamma_{m,p} ^{u -1}(1-\gamma_{m,p} )^{w -1} \\
\end{equation*}
\end{itemize}

\end{document}